\documentclass[preprint]{aastex631}

\usepackage{xspace}
\usepackage{graphicx}
\usepackage{natbib}
\usepackage{url}
\usepackage{bm}
\usepackage{amsmath}
\usepackage[maxfloats=256]{morefloats}

\newcommand{\myr}{\ensuremath{\,{\rm Myr}}}

\newcommand{\ghz}{\ensuremath{\,{\rm GHz}}}
\newcommand{\kel}{\ensuremath{\,{\rm K}}}

\newcommand{\mm}{\ensuremath{\,{\rm mm}}}
\newcommand{\cm}{\ensuremath{\,{\rm cm}}}

\newcommand{\percc}{\ensuremath{\,{\rm cm^{-3}}}}

\newcommand{\pc}{\ensuremath{\,{\rm pc}}}
\newcommand{\kms}{\ensuremath{\,{\rm km\, s^{-1}}}}

\newcommand{\hr}{\,hr}

\newcommand{\EM}{\ensuremath{\,{\rm cm^{-6}\, pc}}}

\newcommand{\expo}[1]{\ensuremath{10^{#1}}}
\newcommand{\nexpo}[2]{\ensuremath{#1 \times 10^{#2}}}

\newcommand{\hii}{H\,{\sc ii}}

\newcommand{\heii}{He\,{\sc ii}}
\newcommand{\oii}{O\,{\sc ii}}
\newcommand{\oiii}{O\,{\sc iii}}

\newcommand{\hepr}[1]{\ensuremath{{}^#1{\rm He}^{+}/{\rm H^+}}}

\newcommand{\mwc}[1]{MWC\thinspace #1}
\newcommand{\sgr}[1]{Sgr\thinspace #1}

\newcommand\urltilda{\kern -.15em\lower .7ex\hbox{\~{}}\kern .04em}

\newcommand{\lsim}{\ensuremath{\lesssim}}

\begin{document}

\title{The Metallicity-Electron Temperature Relationship in \hii\ Regions}

\author[0000-0002-2465-7803]{Dana S. Balser}
\affiliation{National Radio Astronomy Observatory, 520 Edgemont Rd.,
  Charlottesville, VA 22903, USA}

\author[0000-0003-0640-7787]{Trey V. Wenger}
\affiliation{NSF Astronomy \& Astrophysics Postdoctoral Fellow, Department of Astronomy, University of Wisconsin–Madison, Madison, WI 53706, USA}

\begin{abstract}

  \hii\ region heavy-element abundances throughout the Galactic disk
  provide important constraints to theories of the formation and
  evolution of the Milky Way.  In LTE, radio recombination line (RRL)
  and free-free continuum emission are accurate extinction-free
  tracers of the \hii\ region electron temperature. Since metals act
  as coolants in \hii\ regions via the emission of collisionally
  excited lines, the electron temperature is a proxy for
  metallicity. Shaver et al. found a linear relationship between
  metallicity and electron temperature with little scatter.  Here, we
  use CLOUDY \hii\ region simulations to (1) investigate the accuracy
  of using RRLs to measure the electron temperature; and (2) explore
  the metallicity-electron temperature relationship.  We model 135
  \hii\ regions with different ionizing radiation fields, densities,
  and metallicities.  We find that electron temperatures derived under
  the assumption of LTE are about 20\% systematically higher due to
  non-LTE effects, but overall LTE is a good assumption for
  cm-wavelength RRLs.  Our CLOUDY simulations are consistent with the
  Shaver et al. metallicity-electron temperature relationship but
  there is significant scatter since earlier spectral types or higher
  electron densities yield higher electron temperatures.  Using RRLs
  to derive electron temperatures assuming LTE yields errors in the
  predicted metallicity as large as 10\%.  We derive correction
  factors for Log(O/H) + 12 in each CLOUDY simulation.  For lower
  metallicities the correction factor depends primarily on the
  spectral-type of the ionizing star and range from 0.95 to 1.10,
  whereas for higher metallicities the correction factor depends on
  the density and is between 0.97 and 1.05.

\end{abstract}


\section{Introduction}\label{sec:intro}

\hii\ regions are the sites of massive star formation where O and
B-type stars ionize the surrounding gas \citep{stromgren39}.  Since
\hii\ regions are short-lived ($\lsim 10$\myr), their elemental
abundances correspond to present-day values at their location in the
Galaxy.  They therefore provide a snapshot of the distribution of
elemental abundances that are critical to constrain Galactic chemical
evolution models \citep[e.g.,][]{chiappini01}.  \hii\ region abundance
studies complement similar studies of stars, which are typically much
older and have moved from their birthplace
\citep[e.g.,][]{schonrich09}.

\hii\ regions are bright at multiple wavelengths.  They emit copious
amounts of energy via collisionaly excited lines (CELs) from metals at
optical \citep[e.g.,][]{peimbert69} and infrared
\citep[e.g.,][]{simpson95} wavelengths, and they are detected in radio
recombination line (RRL) and continuum emission \citep{hoglund65}.
\hii\ region tracers at radio wavelengths have the advantage that they
are not affected by dust and can be detected throughout the Galactic
disk.  Recently, \hii\ region RRL surveys have nearly tripled the
number of known Galactic \hii\ regions \citep{anderson11, wenger21}.

For an optically thin nebula in local thermodynamic equilibrium (LTE),
the RRL and free-free continuum ratio provides an excellent probe of
the electron (thermal) temperature \citep[e.g.,][]{wilson09}.  In LTE,
both the RRL emission and continuum emission have the same dependence
on electron density, but differ in their dependence on electron
temperature.  Furthermore, since we are forming a line-to-continuum
ratio there is no need for an absolute calibration of the intensity
scale, yielding a very accurate determination of the electron
temperature.  In particular, radio interferometers provide very
precise measurements of the line-to-continuum ratio yielding electron
temperatures with uncertainties of $\sim 1$\% \citep{wenger19b}.
Systematic uncertainties should therefore dominate (e.g., non-LTE
effects).

\citet{churchwell75} were the first to use RRL and continuum data
toward a sample of Galactic \hii\ regions to discover radial electron
temperature gradients in the Milky Way disk \citep[also
  see][]{lichten79}.  They suggested that these results were due to
radial metallicity gradients, similar to those found by observations
of optical CELs in nearby galaxies \citep{searle71}.  Theoretical
models of \hii\ regions predict that the metal abundance is the
dominant factor that determines the electron temperature
\citep[e.g.,][]{rubin85}.  This is because metals act as coolants via
the escape of CEL radiation from primarily oxygen and nitrogen
\citep[e.g.,][]{osterbrock06}.

\citet{shaver83} derived an empirical metallicity-electron temperature
relationship by using optical CELs of oxygen to determine the O/H
abundance ratio, and RRLs to determine the electron temperature,
$T_{\rm e}$, toward a sample of Galactic \hii\ regions.  Since RRLs of
elements heavier than hydrogen and helium are too weak to detect in
\hii\ regions, using the electron temperature as a proxy for
metallicity provides an indirect method to explore metallicity
structure in the Milky Way disk with radio data.  Recent results of
this technique have found azimuthal metallicity structure in the
Galactic disk in addition to the well-established radial metallicity
gradient \citep{balser11, balser15, wenger19b}.

Factors other than metallicity that affect the electron temperature
may not be negligible, however, and should be considered when
evaluating the metallicity-electron temperature relationship.  For
example, the stellar effective temperature, $T_{\rm eff}$, of the
ionizing star determines the hardness of the radiation field that
excites and heats the gas, so higher values of $T_{\rm eff}$ will
marginally increase $T_{\rm e}$ \citep{rubin85}.  The electron
density, $n_{\rm e}$, alters the rate of collisional de-excitation and
so higher values of $n_{\rm e}$ will inhibit cooling and thus slightly
increase $T_{\rm e}$ \citep{rubin85}.  Lastly, dust affects the
electron temperature in complex ways \citep{mathis86, baldwin91,
  shields95}.  Dust can enable heating through the photo-electric
effect as electrons are ejected from dust grains and collide with
atoms.  In contrast, cooling can result when there are collisions of
fast particles with dust grains.

\section{Cloudy Simulations}\label{sec:cloudy}

Here, we investigate the metallicity-electron temperature relationship
by simulating \hii\ regions using the spectral synthesis code CLOUDY
\citep{chatzikos23}.  We model a wide range of \hii\ regions physical
conditions and investigate departures from LTE in RRL emission.  We
use a development version of CLOUDY (trunk branch, revision r13270M)
that includes an update to the energy changing collisonal rates that
is important for predicting RRL emission from ionized gas \citep[for
  details see][]{guzman19}.  For all simulations we assume a spherical
nebula with hydrogen density, $n_{\rm H}$, and diameter, $D$, ionized
by a central star.  Since we are modeling spectral transitions with
high principal quantum number, $n$, a large number of quantum levels
must be considered.  The first 25 levels from hydrogen are resolved
into $nl$ terms, whereas the next 375 levels are collapsed into one
effective level where the $nl$ terms are assumed to be populated
according to their statistical weight.  Here, $l$ is the azimuthal
quantum number related to the angular momentum of the atom.  All
simulations use the semiclassical straight-trajectory Born
approximation of \citet{lebedev98} for the excitation rate
coefficients as recommended by \citet{guzman19}.

We use CLOUDY to explore the wide range of physical properties found
in Galactic \hii\ regions.  Table~\ref{tab:cloudy} summarizes the
\hii\ region properties of the CLOUDY simulations.  We choose three
different spectral types---O3, O6, and O9---in our grid of simulations
with corresponding effective temperatures and number of H-ionizing
photons, $N_{\rm Ly}$, given by \citet{martins05}.  We use the ATLAS
stellar grids which are LTE, plane-parallel, hydrostatic model
atmospheres \citep{castelli03}.  We assume the metallicity of the star
and gas are the same and therefore select the stellar metallicity that
is closest to the gas metallicity (see below).  There are some nearby
\hii\ regions that are ionized by B-type stars, but most \hii\ regions
detected at optical or radio wavelengths require O-type stars
\citep[e.g.,][]{caplan00, bania07}.

\citet{kurtz05} classifies \hii\ regions based on their density and
size.  We approximately follow this classification in our grid of
simulations by modeling ultra-compact to giant \hii\ regions.  This
sets the total hydrogen density and size of the model nebula where
more compact sources will have higher densities.  The range of
($n_{\rm H}, D$) pairs corresponds to emission measures $EM \equiv
\int{n_{\rm e}^{2} d\ell} = 10^{3}-10^{7}$\EM\ for a fully ionized
nebula.  We therefore call these ($n_{\rm H}, D$) pairs EM3, EM4,
etc. (see Table~\ref{tab:cloudy}).

The nominal elemental abundance ratios relative to hydrogen are
specified by the CLOUDY \hii\ region model for Orion.  The abundances
are determined by calculating the mean value from three independent
studies.  We vary the metallicity via the CLOUDY {\tt metals} command
which scales the abundance ratios for all elements heavier than
helium.  This corresponds to O/H abundance ratios between Log(O/H) +
12 = 7.8 and 9.4 that conservatively encompasses the values determined
across the Milky Way disk \citep[e.g.,][]{deharveng00, rudolph06,
  arellano-cordova20, arellano-cordova21}.  We do not include dust for
two reasons.  First, \citet{oliveira86} estimate that the net effect
of dust on the electron temperature is relatively small.  Second, for
some simulations that included Orion dust grains, we received warnings
that the dust would not survive at these temperatures and thus the
simulations were not realistic.

\begin{deluxetable}{cccrrcccc} 
\tablecolumns{9} \tabletypesize{\small} \tablecaption{CLOUDY
  Simulation \hii\ Region Properties \label{tab:cloudy}} \tablehead{
  \multicolumn{3}{c}{Ionizing Star} & \colhead{} &
  \multicolumn{3}{c}{Cloud Properties} & \colhead{} &
  \colhead{Metallicity} \\ \cline{1-3} \cline{5-7} \cline{9-9}
  \\ \colhead{Spectral} & \colhead{$T_{\rm eff}$} &
  \colhead{Log($N_{\rm Ly}$)} & \colhead{} & \colhead{$n_{\rm H}$} &
  \colhead{$D$} & \colhead{Name} & \colhead{} & \colhead{Log(O/H) +
    12} \\ \colhead{Type} & \colhead{(K)} & \colhead{(s$^{-1}$)} &
  \colhead{} & \colhead{(cm$^{-3}$)} & \colhead{(pc)} & \colhead{} &
  \colhead{} & \colhead{} } \startdata
O3      & 44,616   & 49.63 & &       10   & 10.0    & EM3 & & 7.8 \\
O6      & 38,151   & 48.63 & &       50   &  4.0    & EM4 & & 8.0 \\
O9      & 31,524   & 47.90 & &      200   &  2.5    & EM5 & & 8.2 \\
\nodata & \nodata  & \nodata & &  1,000   &  1.0    & EM6 & & 8.4 \\
\nodata & \nodata  & \nodata & & 10,000   &  0.1    & EM7 & & 8.6 \\
\nodata & \nodata  & \nodata & & \nodata  & \nodata & \nodata  & & 8.8 \\
\nodata & \nodata  & \nodata & & \nodata  & \nodata & \nodata  & & 9.0 \\
\nodata & \nodata  & \nodata & & \nodata  & \nodata & \nodata  & & 9.2 \\
\nodata & \nodata  & \nodata & & \nodata  & \nodata & \nodata  & & 9.4 \\
\enddata
\tablecomments{The CLOUDY simulations consider 3 different spectral
  types with corresponding effective temperature ($T_{\rm eff}$) and
  hydrogen ionizing photon rates ($N_{\rm Ly}$); 5 different spherical
  nebulae with density ($n_{\rm H}$) and diameter ($D$); and 9
  different metallicities characterized by the O/H abundance ratio.
  This corresponds to $3 \times 5 \times 9 = 135$ simulations.}
\end{deluxetable}

What is the optimal RRL transition to observe when using radio data to
determine the electron temperature?  To derive accurate electron
temperatures requires (1) that the RRL and free-free continuum
emission be optically thin; and (2) that the RRL be formed in LTE.  To
accommodate (1) in classical \hii\ regions the RRL frequency must be
greater than about 5\ghz\ or Hn$\alpha$ transitions with $n \lsim 109$
\citep[e.g.,][]{wilson09}.  RRL emission from most Galactic
\hii\ regions is typically close to LTE.  This is because the physical
conditions necessary for non-LTE effects, low electron densities and
high emission measure, are not common in Galactic \hii\ regions
\citep[see][]{shaver80a}.  This depends, however, on the detailed
physical conditions, geometry, and RRL frequency.  For example,
significant stimulated emission was found toward \mwc349, which may
comprise of a rotating and expanding disk, from the H30$\alpha$
transition at 1\mm\ \citep{martin-pintado89}.  Pressure broadening
from electron impacts that decrease the peak line intensity was
detected toward \sgr{B2}, which contains many high emission measure
components, from the H109$\alpha$ transition at
6\cm\ \citep{vonprochazka10}.  \citet{shaver80b} has determined that
the optimal frequency to observe RRLs such that the electron
temperature derived assuming LTE is equal to the true electron
temperature is given by:
\begin{equation}
  \nu \sim 0.081\,EM^{0.36}\,\,\, {\rm GHz}.
  \label{eq:em}
\end{equation}
For typical Galactic \hii\ region emission measures this corresponds
to cm-wavelength RRLs.  This is, in part, why recent studies of
Galactic metallicity structure have used cm-wavelength RRLs
\citep{wenger19b}.  Here, we therefore focus on the H87$\alpha$
transition at 3.05\cm\ as a representative RRL.

In total there are 135 CLOUDY simulations which correspond to 3
spectral types $\times$ 5 emission measures $\times$ 9 metallicities.
Appendix~\ref{sec:sim} includes the input parameters for one CLOUDY
simulation as an example.  The simulations were run on an Intel Xeon
Silver 4114 processor at 2.20\ghz.  We only used 1 of the available 20
cores since the total memory (RAM) was only 251.4$\,$GB and each
simulation required $\sim 100\,$GB.  Because of the large number of
quantum levels that were considered each simulation took about
2.5\hr\ to run.

\section{Results}\label{sec:results}

All 135 CLOUDY simulations exited without errors after two iterations.
The nebula was typically divided into about 200 numerical zones.  For
some simulations with high metallicities the nebular electron
temperature fell below 4,000\kel, the default threshold value below
which the CLOUDY calculation will be stopped.  This is the default
value because thermal instabilities may occur when the temperature is
below $\sim \nexpo{\rm few}{3}$\kel\ since the cooling curve allows
more than one thermal solution \citep{williams67}.  We therefore reran
all simulations with a threshold temperature of 1,000\kel\ to allow
for more realistic electron temperatures.  For 30 simulations we
lowered the threshold temperature to 500\kel\ because the temperature
dropped below 1,000\kel\ halting the simulation.  We received no
warnings from CLOUDY that the heating-cooling balance was not
preserved in these cases and therefore the temperatures should be
reliable.

\begin{figure}
  \centering
  \includegraphics[angle=0,scale=0.55]{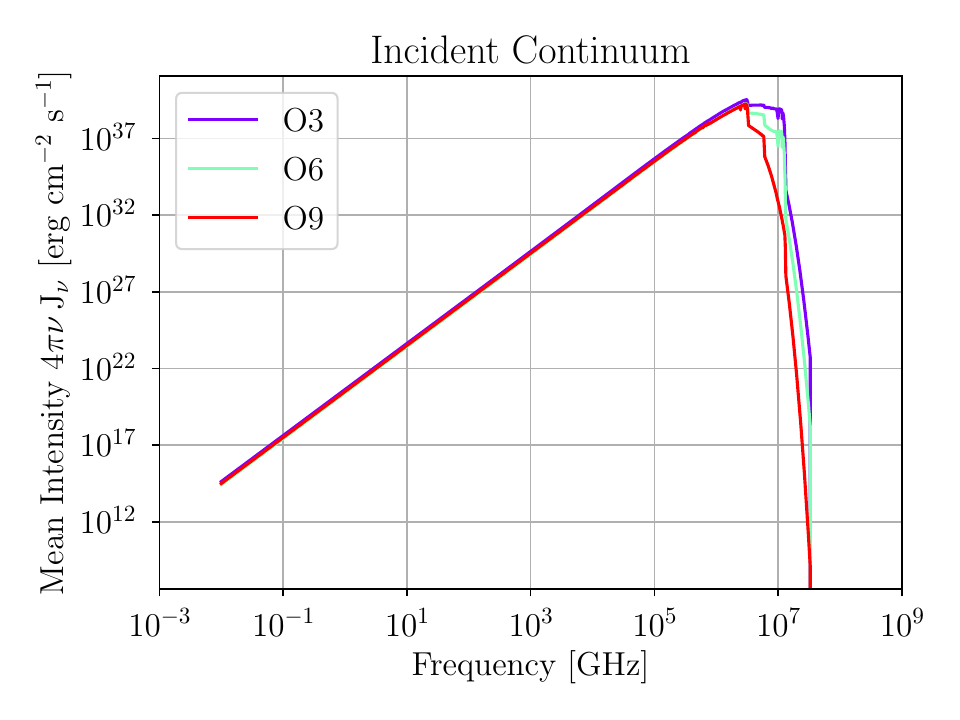}
  \includegraphics[angle=0,scale=0.55]{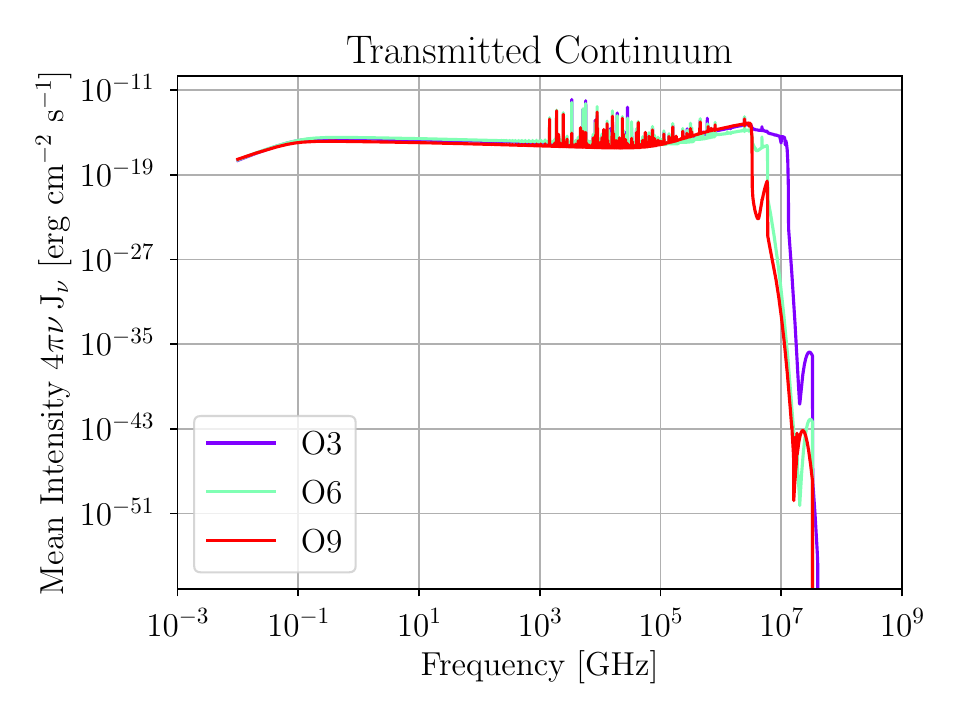}
  \caption{Incident and transmitted continuum spectra for three CLOUDY
    simulations.  For all three simulations: $n_{\rm H} = 200$\percc,
    $D = 2.5$\pc, and Log(O/H) + 12 = 8.6.  This corresponds to an
    emission measure of $EM = \expo{5}$\EM\ for a fully ionized
    nebula.  Each plot shows the results for ionizing stars with
    spectral-types O3, O6, and O9.  The transmitted continuum emission
    includes the spectral lines which are dominated by infrared and
    optical collisionally exited lines.  At lower frequencies RRLs are
    present but not visible on this scale.}
  \label{fig:cont}
\end{figure}

Figure~\ref{fig:cont} shows the incident and transmitted continuum for
three representative simulations.  As expected the simulations with
earlier spectral type stars have harder incident radiation fields.
The transmitted continuum includes emission from spectral lines.
Notable are the CELs at infrared and optical frequencies.  The RRLs
are present but not visible on this scale.

\begin{figure}
  \centering
  \includegraphics[angle=0,scale=0.55]{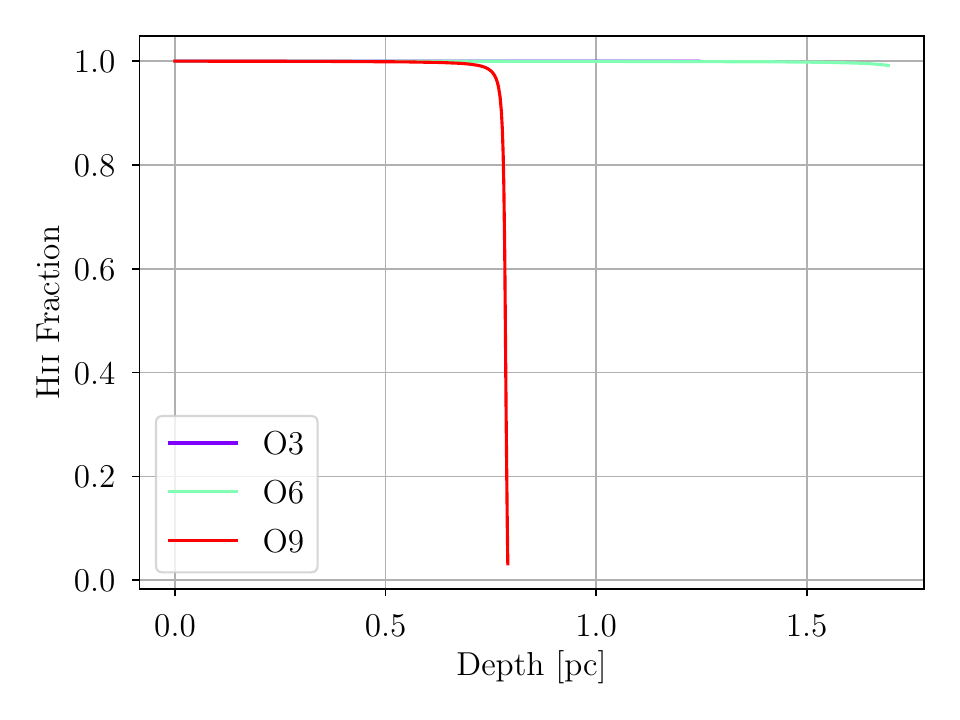}
  \includegraphics[angle=0,scale=0.55]{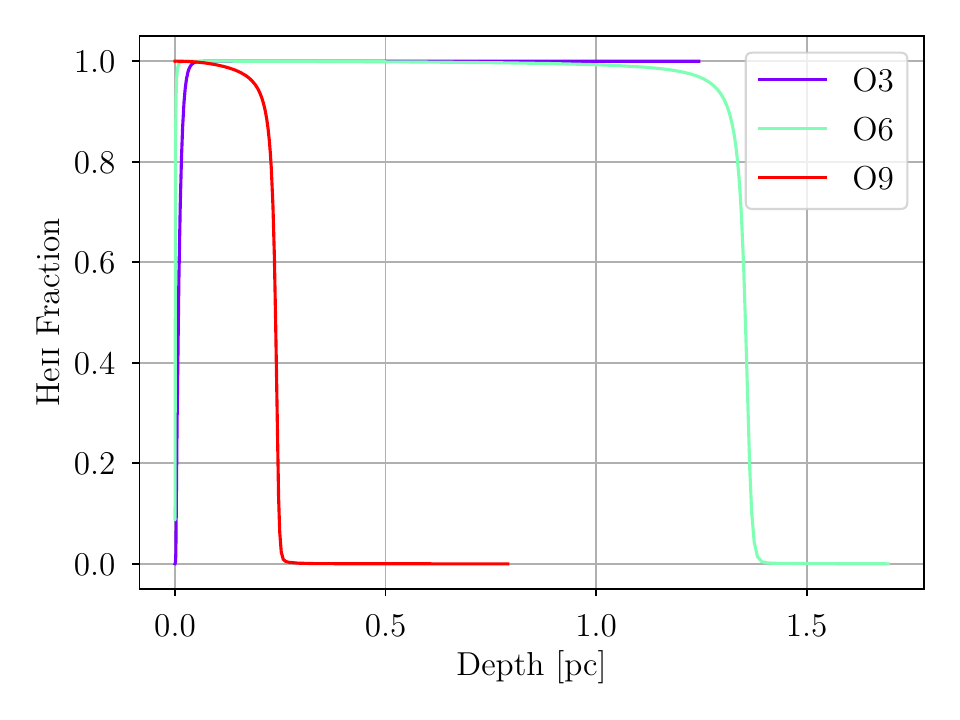}
  \includegraphics[angle=0,scale=0.55]{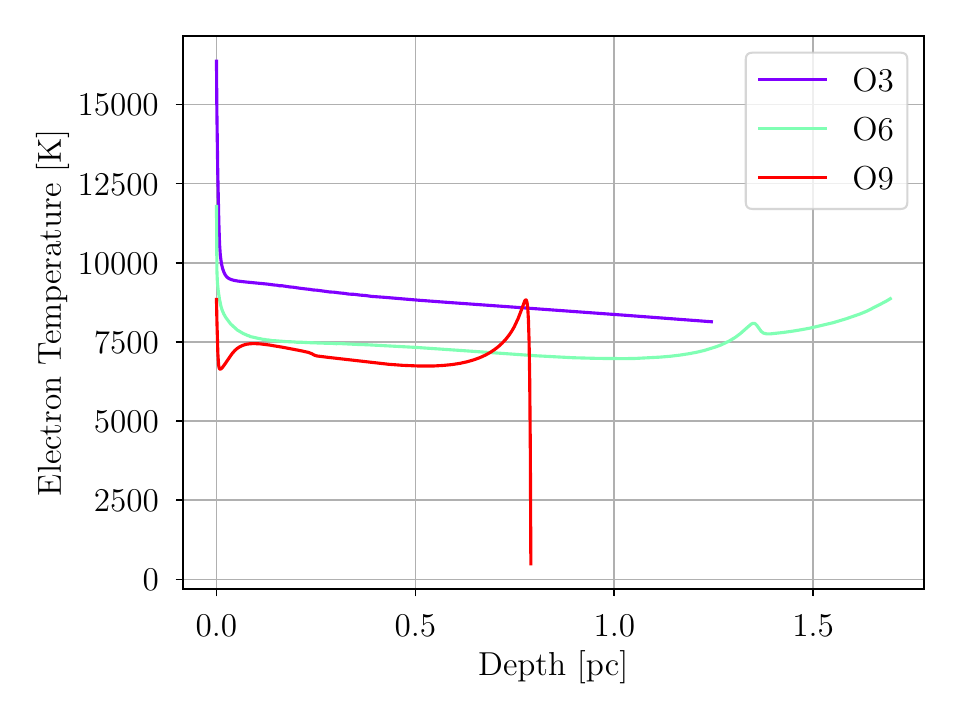}
  \includegraphics[angle=0,scale=0.55]{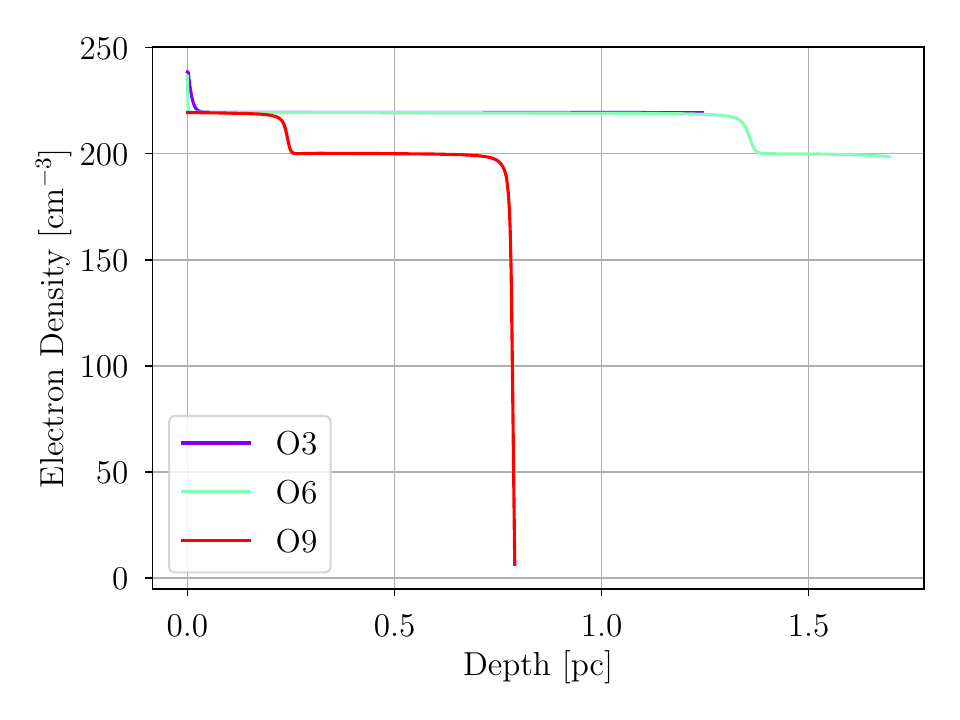}
  \caption{Physical properties as a function of nebula depth for three
    CLOUDY simulations.  For all three simulations: $n_{\rm H} =
    200$\percc, $D = 2.5$\pc, and Log(O/H) + 12 = 8.6.  This
    corresponds to an emission measure of $EM = \expo{5}$\EM\ for a
    fully ionized nebula.  Shown are the ionization fraction of
    \hii\ and \heii\ (top) and the electron temperature and density
    (bottom).  Each plot show the results for ionizing stars with
    spectral-types O3, O6, and O9.  The simulations with
    spectral-types O3 and O6 fully ionize both H and He, whereas the
    simulation with spectral-type O9 is ionization bounded.}
  \label{fig:vs_depth}
\end{figure}

Figure~\ref{fig:vs_depth} illustrates how several physical properties
vary with depth into the nebula for the same three simulations as
shown in Figure~\ref{fig:cont}.  Included are the \hii\ and
\heii\ ionization fraction and the electron temperature and density.
The nebulae ionized by either O3 or O6-type stars are density-bounded
where the hydrogen and helium are fully ionized.  In contrast, the
nebula ionized by the O9 star is ionization-bounded where neutral gas
lies beyond the ionized gas.  The electron temperature and density are
relatively constant with depth but there are some variations.  For
example, the electron temperature slightly increases at the
\hii\ region boundary because of {\it photon hardening}
\citep[e.g.,][]{osterbrock06, wilson15}.  The electron density
decreases past the \heii\ region boundary since fewer electrons are
available from helium.

Since our goal is to use the electron temperature as a proxy for
metallicity, we need to determine a representative value for the
entire nebula.  First, we derive a ``real'' electron temperature using
values within each numerical zone.  Specifically, we calculate a
$n_{\rm e}n_{\rm p}$-weighted value averaged over the volume of the
model spherical nebula, denoted as $T_{\rm e}^{n_{\rm e}n_{\rm p}}$.
Here, $n_{\rm p}$ is the proton density.  This weighting is
appropriate for radio studies with tracers that depend on the emission
measure.  We also derive a ``synthetic'' electron temperature based on
observable radio diagnostics produced by the CLOUDY simulations.  We
use the synthetic H87$\alpha$ RRL and free-free continuum intensity
that escapes the model nebula to derive the LTE electron temperature,
$T_{\rm e}^{*}$, given by
\begin{equation}
  T_{\rm e}^{*} = \Big[ \nexpo{7.100}{3} \Big( \frac{I_{\rm C}(\nu_{\rm L})}{I_{\rm L}(\nu_{\rm L})} \Big)
    \Big( \frac{\nu_{\rm L}}{\rm GHz} \Big)^{1.1} \Big( \frac{\Delta V}{\kms} \Big)^{-1} (1 + y^{+})^{-1}
    \Big]^{0.87}\kel,
\label{eq:LTE}
\end{equation}
where $I_{\rm L}(\nu_{\rm L})$ and $I_{\rm C}(\nu_{\rm L})$ are the
RRL and free-free continuum intensities, respectively, at the RRL
frequency $\nu_{\rm L}$, $\Delta V$ is the full-width at half-maximum
RRL width, and $y^{+}$ is the singly ionized helium abundance ratio,
\hepr4\ \citep[see][]{wenger19b}.

\subsection{Departures from LTE}\label{sec:LTE}

\begin{figure}
  \centering
  \includegraphics[angle=0,scale=0.55]{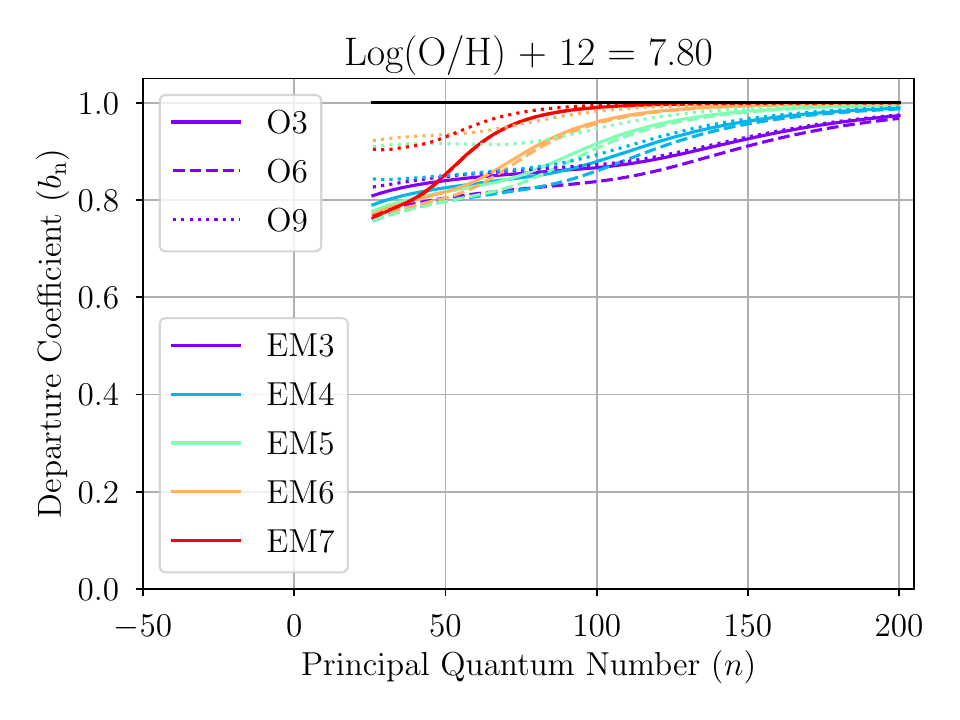}
  \includegraphics[angle=0,scale=0.55]{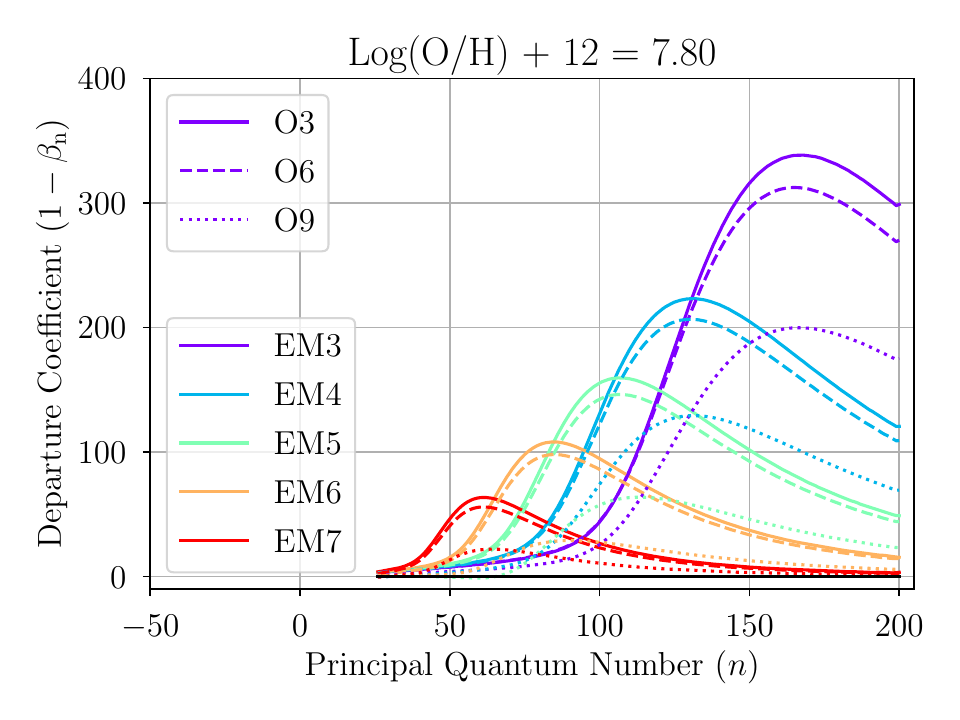}
  \includegraphics[angle=0,scale=0.55]{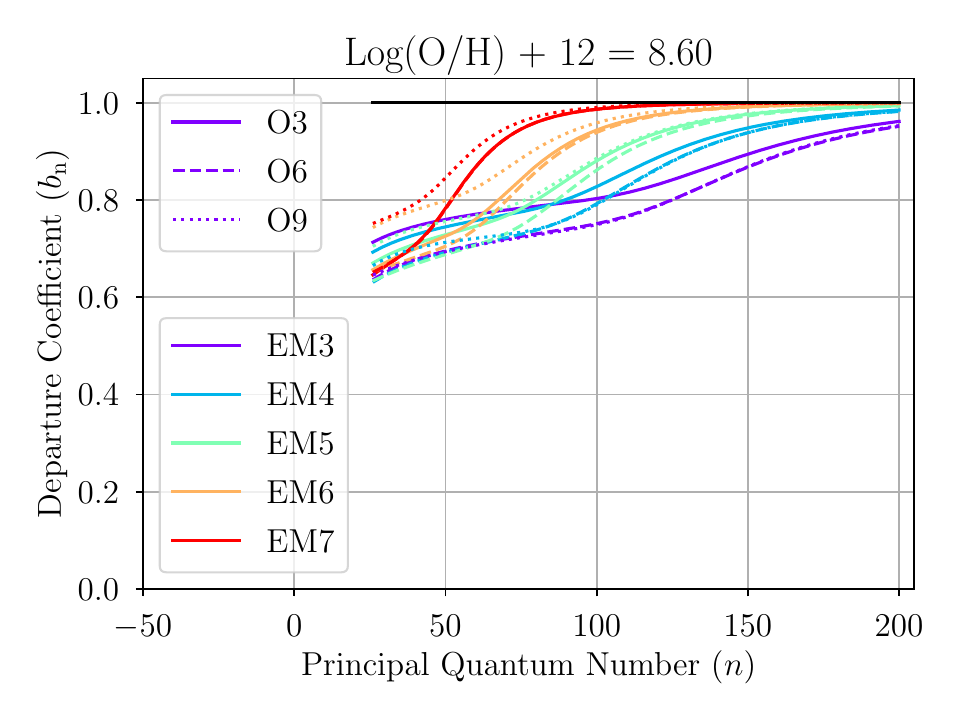}
  \includegraphics[angle=0,scale=0.55]{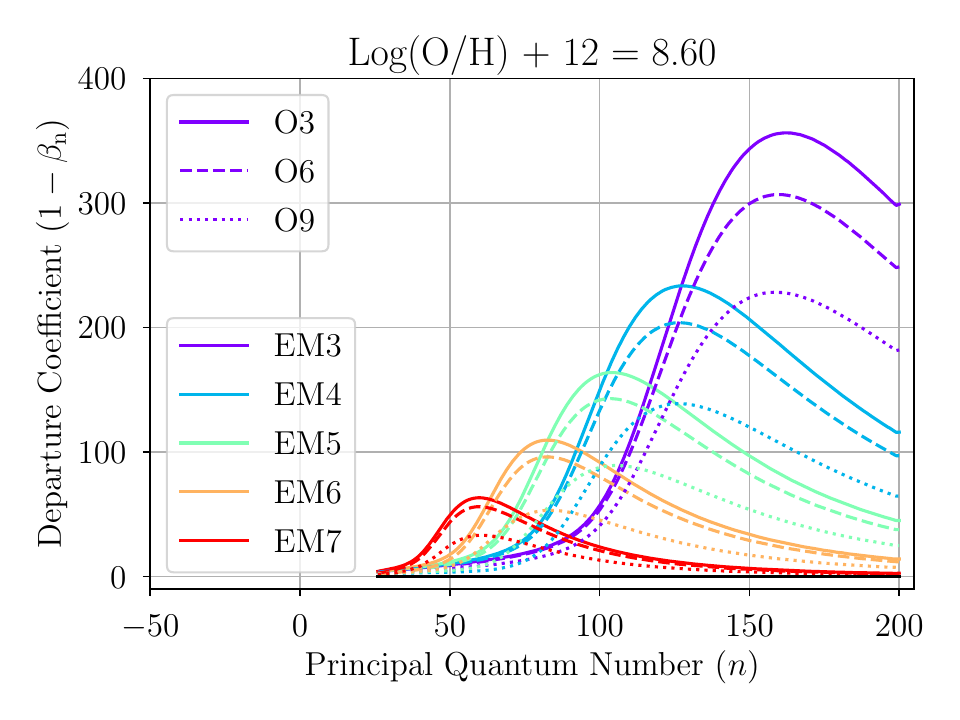}
  \includegraphics[angle=0,scale=0.55]{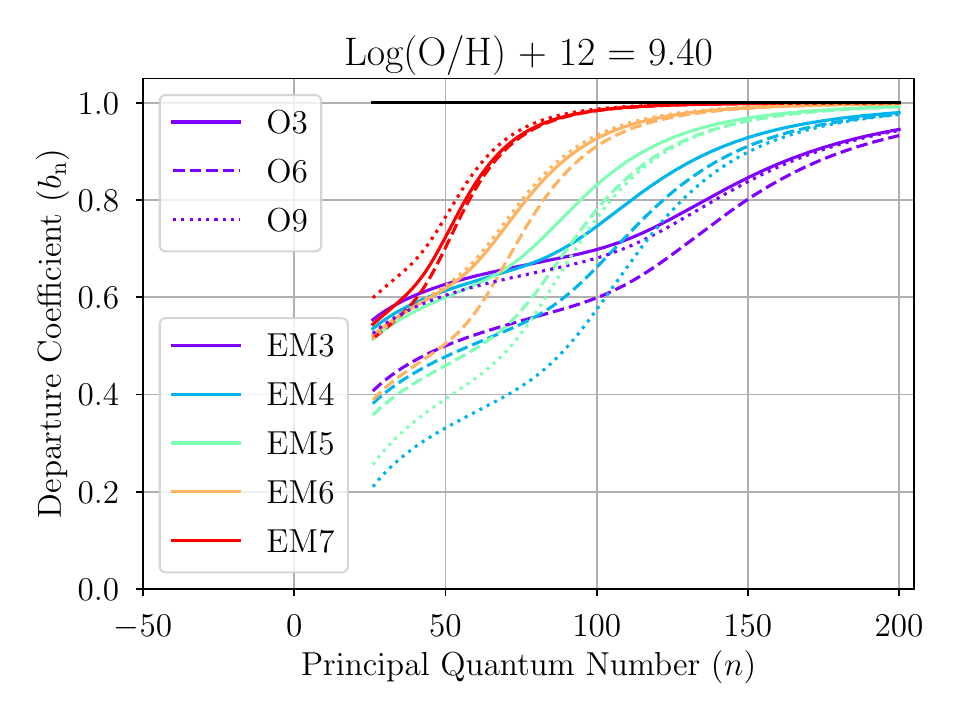}
  \includegraphics[angle=0,scale=0.55]{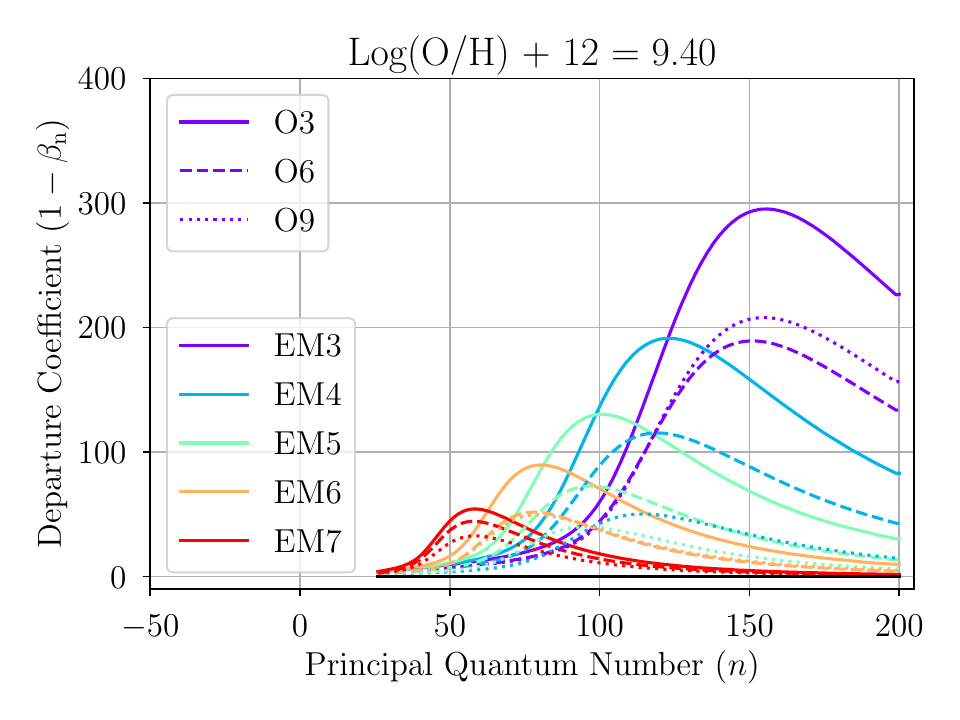}
  \caption{Departures from LTE in our CLOUDY simulations.  Plotted are
    the departure coefficients $b_{\rm n}$ (left) and $\beta_{\rm n}$
    (right) as a function of $n$.  Different metallicities are shown
    with Log(O/H) + 12 values of 7.8 (top), 8.6 (middle), and 9.4
    (bottom).  The line color represents different emission measures,
    whereas the line style corresponds to different spectral types.
    The departure coefficients were taken from a representative
    numerical cell that has a depth 25\% into the \hii\ region from
    the center of the model nebula.  For cm-wavelength RRLs, $n \sim
    100$.}
  \label{fig:non-LTE}
\end{figure}

The advantage of assuming LTE is that the electron temperature can be
derived independent of the geometry and density structure of the
\hii\ region which are difficult to constrain with observations.
Here, we consider departures from LTE to assess the limitations of
using RRLs as a diagnostic for metallicity.  Figure~\ref{fig:non-LTE}
summarizes non-LTE effects for our CLOUDY simulations.  Plotted are
the departure coefficients $b_{\rm n}$ (left) and $\beta_{\rm n}$
(right) as a function of $n$.  The departure coefficients depend on
the physical properties of the ionized gas, primarily the electron
density and temperature, for a given $n$ \citep[e.g.,
  see][]{brocklehurst72}.

The departure coefficient $b_{\rm n}$ is given by
\begin{equation}
  b_{\rm n} = N_{\rm n}/N_{\rm n}^{*},
\label{eq:bn}
\end{equation}
where $N_{\rm n}$ is the true population level and $N_{\rm n}^{*}$ is
the population level in LTE for quantum level $n$.  Since the Einstein
$A$ coefficient for the lower state is larger and the atom is smaller
so that collisions are less effective, $b_{\rm n} < 1$.  For $n =
100$, $b_{\rm n} > 0.8$ for most simulations but larger departures
from LTE exist for the higher metallicities where the electron
temperatures are smaller.  Simulations with lower electron densities
have larger departures from LTE since there are fewer collisions.

The departure coefficient $\beta_{\rm n}$ is a measure of the gradient
of $b_{\rm n}$ with respect to $n$ and is given by \citep{wilson09}:
\begin{equation}
  \beta_{\rm n} = 1 - 20.836 \Big( \frac{T_{\rm e}}{\rm K} \Big)
  \Big( \frac{\nu_{\rm L}}{\rm GHz} \Big)^{-1} \frac{{\rm d}\,{\rm ln}\, b_{\rm n}}{{\rm d}n} \Delta{n}.
\label{eq:beta}
\end{equation}
The values of $\beta_{\rm n}$ can be much different than unity and for
inverted populations, $\beta_{\rm n} < 0$, maser amplification occurs.
For our simulations we see values as high as $\beta_{\rm n} \sim
-300$, but how this alters the RRL intensities will depend on the
detailed radiative transfer and continuum opacity \citep[e.g.,
  see][]{wilson09}.

How do the departure coefficients affect the electron temperature?
Following \citet{wilson09}, we estimate the non-LTE RRL electron
temperature as
\begin{equation}
  T_{\rm e} = T_{\rm e}^{*}\,[b_{\rm n}(1 - \frac{1}{2} \beta_{\rm n} \tau_{\rm c})]^{0.87},
\label{eq:non-LTE}
\end{equation}
where $\tau_{\rm c}$ is the free-free continuum opacity.
Equation~\ref{eq:non-LTE} is an approximation to a uniform region with
a background opacity $\tau_{\rm c}$.  In practice, the equation of
transfer must be solved from the back of the nebula to the front with
respect to the observer.  If the nebula is optically thin, however,
($1 - \frac{1}{2} \beta_{\rm n} \tau_{\rm c}) \sim 1$ and so $T_{\rm
  e} \sim T_{\rm e}^{*}\,b_{\rm n}^{0.87}$.  Because most nebulae are
optically thin at cm-wavelengths, deviations from LTE will be mostly
due to $b_{\rm n}$.  Since $b_{\rm n} < 1$, electron temperatures
calculated assuming LTE will overestimate the true electron
temperature.

\begin{figure}
  \centering
  \includegraphics[angle=0,scale=0.55]{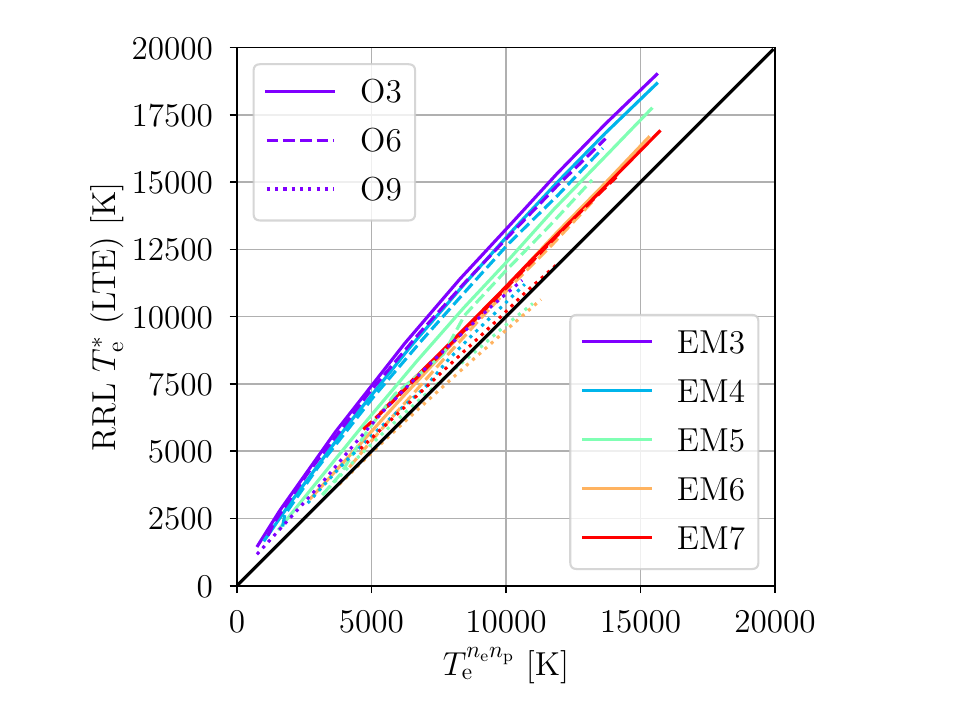}
  \includegraphics[angle=0,scale=0.55]{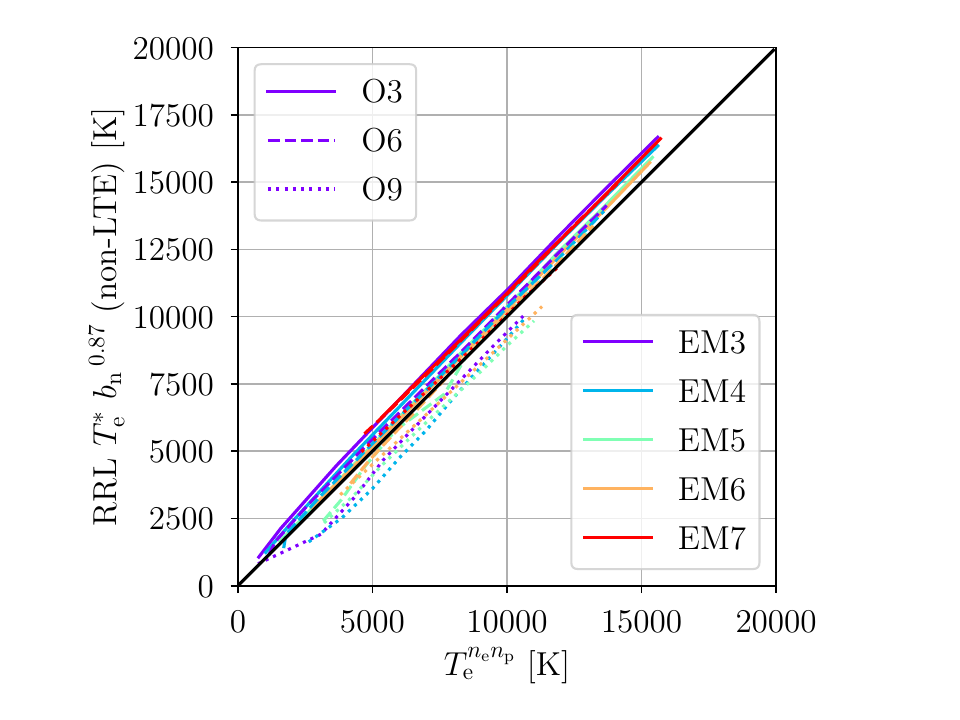}
  \includegraphics[angle=0,scale=0.42]{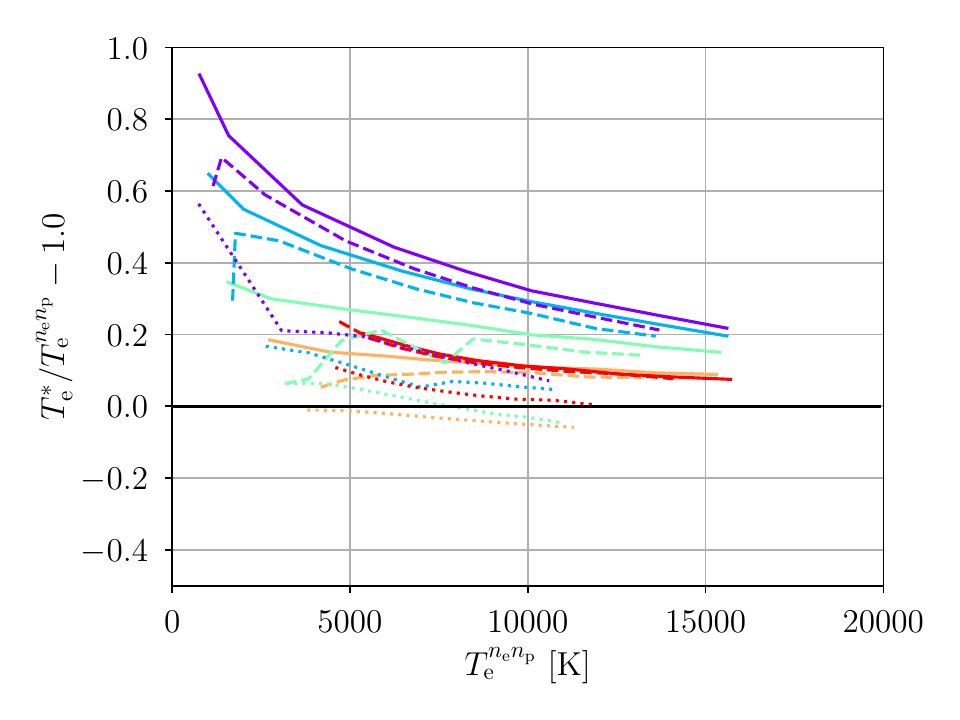}
  \hspace{2.2cm}\includegraphics[angle=0,scale=0.42]{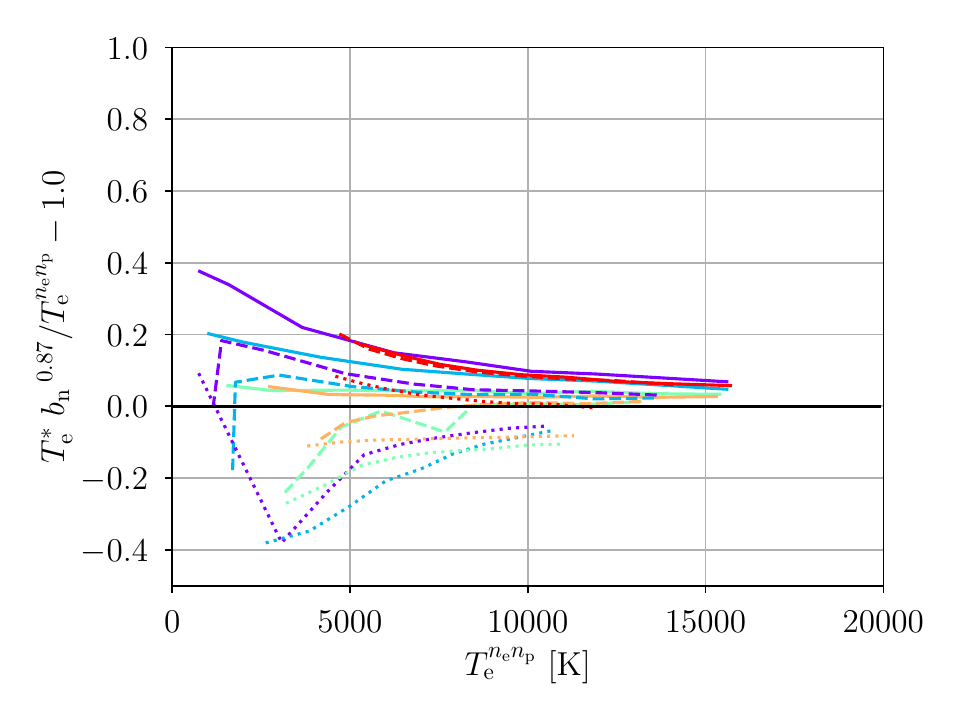}
  \caption{Accuracy of electron temperatures derived using RRLs.  The
    electron temperature, $T_{\rm e}^{n_{\rm e}n_{\rm p}}$, is
    determined by averaging the $n_{\rm e}n_{\rm p}$-weighted electron
    temperatures over the volume of the model spherical nebula.  The
    RRL-derived electron temperature is calculated from the synthetic
    H87$\alpha$ RRL and free-continuum emission produced by the CLOUDY
    simulation assuming LTE (left) and non-LTE (right).  Here we
    assume the gas is optically thin so $T_{\rm e} = T_{\rm
      e}^{*}\,b_{\rm n}^{0.87}$ \citep{wilson09}.  We plot the results
    of this comparison for all 135 simulations.  The nine different
    metallicities provide the wide range of electron temperatures
    shown as curves in the plots.  The line color represents different
    emission measures, whereas the line style corresponds to different
    spectral types.  The bottom panels plot the fractional difference
    to reveal details.}
  \label{fig:te}
\end{figure}

Figure~\ref{fig:te} compares the ``real'' electron temperature,
$T_{\rm e}^{n_{\rm e}n_{\rm p}}$, to the ``synthetic'' electron
temperature derived using RRLs assuming LTE, $T_{\rm e}^{*}$, for all
135 simulation.  As expected, $T_{\rm e}^{*} > T_{\rm e}^{n_{\rm
    e}n_{\rm p}}$ for almost all simulations.  Electron temperatures
derived assuming LTE are about 20\% systematically higher than the
``real'' electron temperature.  For model nebula with low densities
and high metallicities the difference can be as high as 50\%.  The
plots on the right correct the electron temperatures for departures
from LTE assuming an optically thin nebula and account for most, but
not all, of the discrepancy.

\subsection{CLOUDY Metallicity-Electron Temperature Relationship}\label{sec:relationship}

\begin{figure}
  \centering
  \includegraphics[angle=0,scale=0.55]{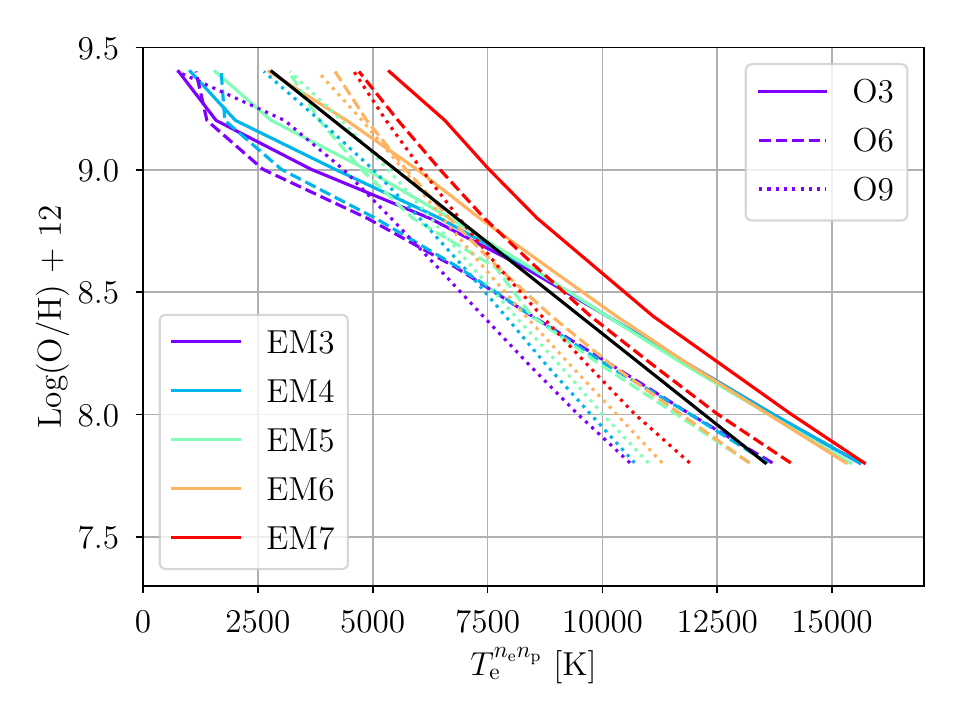}
  \includegraphics[angle=0,scale=0.55]{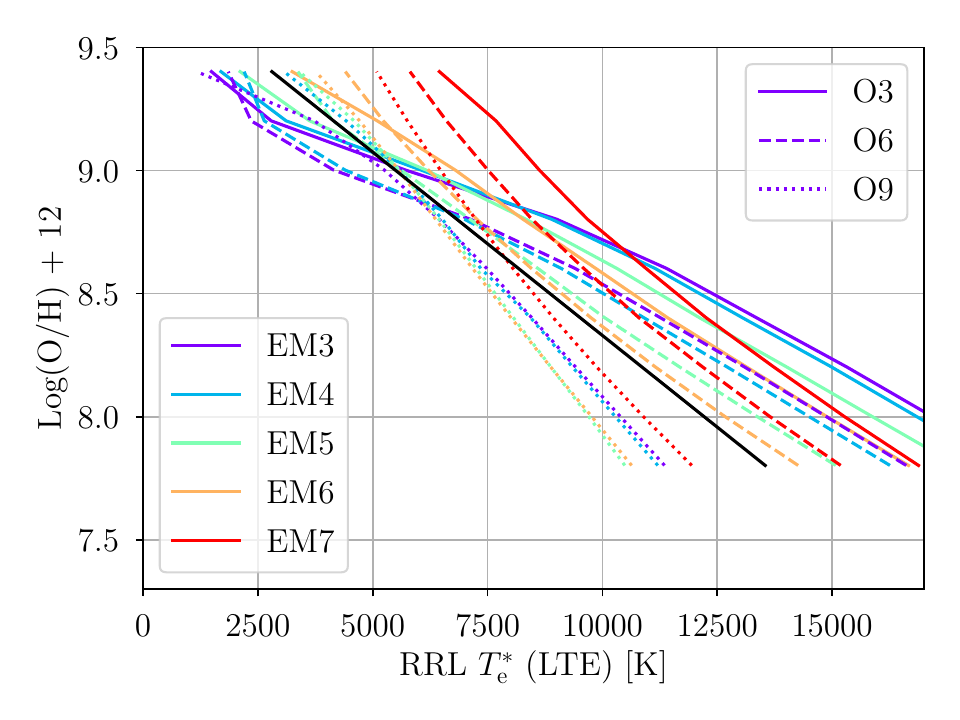}
  \caption{The oxygen abundance ratio, Log(O/H) + 12, as a function of
    electron temperature for all 135 CLOUDY simulations.  {\it Left:}
    The electron temperature is determined by averaging the $n_{\rm
      e}n_{\rm p}$-weighted electron temperatures over the volume of
    the model spherical nebula. {\it Right:} The RRL-derived electron
    temperature calculated from the synthetic H87$\alpha$ RRL and
    free-continuum emission produced by the CLOUDY simulation assuming
    LTE.  The line color represents different emission measures and
    the line style corresponds to different spectral-types.  The black
    solid line is the metallicity-electron temperature relationship
    from \citet{shaver83}.  Generally, the electron temperature
    increases with earlier spectral-type stars and increasing electron
    density.}
  \label{fig:te-o2h}
\end{figure}

There is an empirical correlation between the metallicity, probed by
the O/H abundance ratio, and electron temperature because the
metallicity is the main factor that regulates the temperature in
\hii\ regions \citep[e.g.,][]{rubin85}.  But the ionizing radiation
field spectrum and the nebular density also influence the thermal
properties of the \hii\ region.  Figure~\ref{fig:te-o2h} summarizes
the metallicity-electron temperature relationship produced by the
CLOUDY simulations.  The left panel plots Log(O/H) + 12 as a function
of the ``real'' electron temperature, $T_{\rm e}^{n_{\rm e}n_{\rm
    p}}$, for all 135 simulations.  The line color represents
different emission measures, whereas the line style corresponds to
different spectral types.  The black solid line is the empirical
metallicity-electron temperature relationship determined by
\citet{shaver83} and given by:
\begin{equation}
{\rm Log(O/H)} + 12 = (9.82 \pm 0.02) - (1.49 \pm 0.11)T_{\rm e}/10^{4}.
\label{eq:shaver}
\end{equation}
This empirical relationship is consistent with the CLOUDY results but
there is considerable scatter due to differences in the spectral type
of the ionizing star and the nebular electron density.  For example,
simulations with earlier type stars or higher electron densities have
higher electron temperatures.  The right panel plots the same
relationship using the ``synthetic'' electron temperature, $T_{\rm
  e}^{*}$.  Here the scatter is larger with a systematic offset.  This
is primarily due to departures from LTE that can overestimate the
electron temperature, especially for earlier type stars with lower
emission measures (see Section~\ref{sec:LTE}).

\section{Discussion}\label{sec:discussion}

To determine the Galactic metallicity structure from radio data
requires RRL and free-free continuum observations in \hii\ regions and
a metallicity-electron temperature relationship
\citep[e.g.,][]{wenger19b}.  Here we show that non-LTE effects and
variations in the physical properties of the ionized gas produce
systematic errors in the Galactic metallicity structure.  RRL and
free-free continuum emission toward \hii\ regions provide an accurate
measure of $T_{\rm e}$ if the nebula is optically thin and the ionized
gas is in LTE.  Studies have shown that at cm-wavelengths that these
conditions are valid in many \hii\ regions \citep{shaver80a,
  shaver80b}.  Pressure broadening via electron impacts can alter the
line shape, causing an underestimate of the integrated RRL emission
when the spectral baselines are not well behaved.  This can especially
be an issue for single-dish radio telescopes \citep[e.g.,
  see][]{balser99}, but is not a significant issue for radio
interferometers \citep[e.g.,][]{balser22}.  Stimulated emission is not
common at cm-wavelengths and typically requires a bright background
source with a specific geometry \citep[e.g.,][]{martin-pintado89}.

\begin{figure}
  \centering
  \includegraphics[angle=0,scale=0.55]{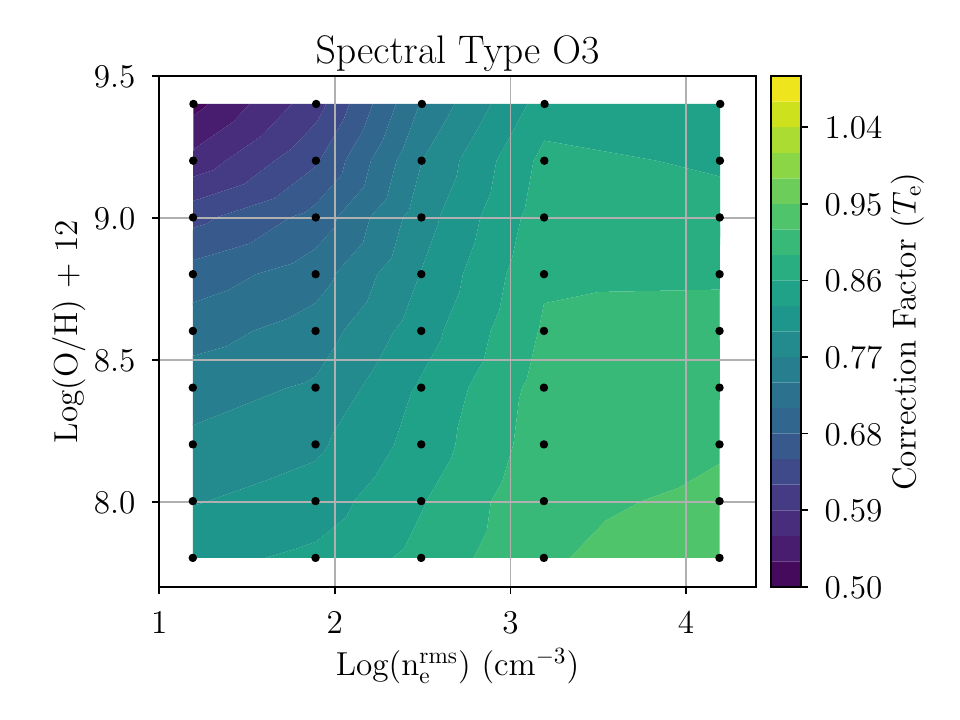}
  \includegraphics[angle=0,scale=0.55]{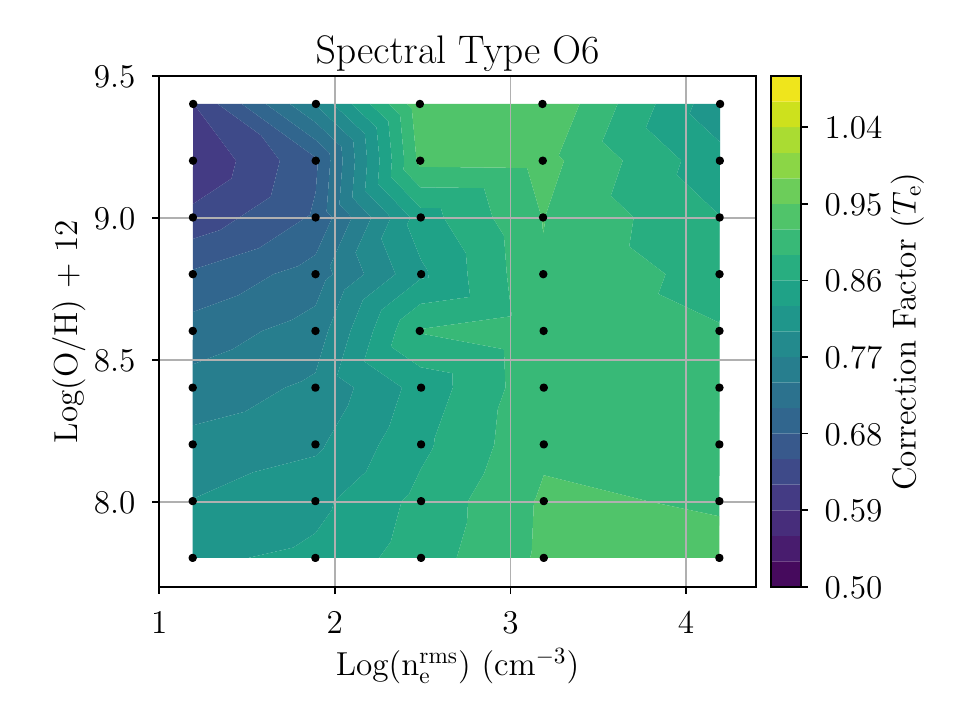}
  \includegraphics[angle=0,scale=0.55]{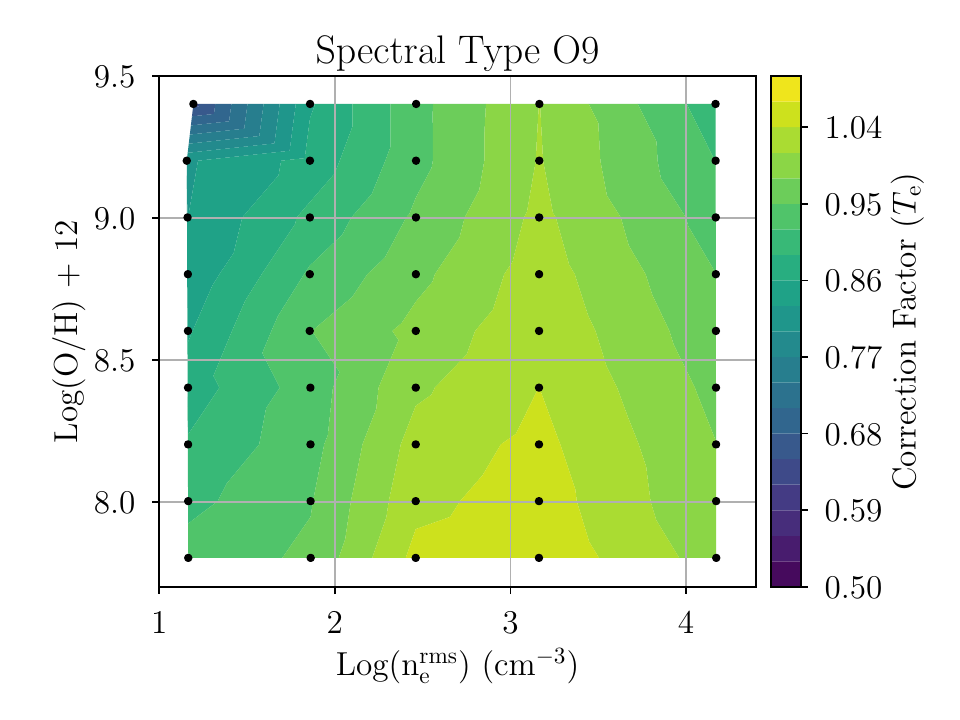}
  \caption{Correction factors of the electron temperature for ionizing
    stars with spectral-types O3, O6, and O9.  The correction factor
    is the ratio of the ``real'' electron temperature to the
    ``synthetic'' electron temperature, $T_{\rm e}^{n_{\rm e}n_{\rm
        p}}$/$T_{\rm e}^{*}$.  The black points correspond to the
    location, defined by Log(O/H) + 12 and $n_{\rm e}^{\rm rms}$ from
    each simulation, from which the contours are generated.  The
    contour levels are between 0.5 and 1.1 with an increment of 0.03.
    For most simulations the predicted electron temperature is
    overestimated yielding correction factors less than unity.}
  \label{fig:cfte}
\end{figure}

Our CLOUDY simulations show, however, that non-LTE effects are
important when attempting to derive {\it accurate} electron
temperatures.  In particular, electron temperatures are systematically
higher by about 20\% when assuming LTE, and in some cases are 50\%
larger for low density nebula (see Figure~\ref{fig:te}).  Since we are
mostly interested in metallicity structure (e.g., radial or azimuthal
gradients), systematic offsets are less important than the dispersion
in these offsets.  Nevertheless, such uncertainties need to be
included in any analysis of metallicity structure when using RRL and
continuum data to derive electron temperatures.  We therefore
calculate correction factors for the electron temperature when LTE is
assumed.  Specifically, the correction factor is the ratio of the
``real'' electron temperature to the ``synthetic'' electron
temperature, $T_{\rm e}^{n_{\rm e}n_{\rm p}}$/$T_{\rm e}^{*}$.  The
electron temperature correction factors are shown as contour plots in
Figure~\ref{fig:cfte} for spectral types O3, O6, and O9.  The
correction factors are typically less than unity for most simulations
since the electron temperatures are systematically higher when
assuming LTE.  For nebulae ionized by O9 stars, however, the
correction factors are closer to unity.

The metallicity-electron relationship exists since heavy elements
primarily regulate the thermal properties of the ionized gas.  To
determine the metallicity-electron temperature \hii\ region
relationship requires the abundance of a heavy element (e.g., oxygen)
relative to hydrogen.  Many studies employ CELs at optical wavelengths
since they are bright.  To derive O/H directly, however, requires the
electron temperature---the $T_{\rm e}$-method.  Electron temperatures
are often determined using the ratio of nebular lines to the higher
energy level auroral lines; for example, [\oiii]$\lambda\lambda$4959,
5007/$\lambda$4363 \citep[e.g.,][]{peimbert69}.  Unfortunately,
[\oiii]$\lambda$4363 is weak and often not detected, so indirect
methods have been developed.  For example, \citet{pagel79} suggested
the $R_{23}$-method, an empirical strong line method, which uses
bright transitions to form the ratio ([\oii]$\lambda\lambda$3726, 3729
+ [\oiii]$\lambda\lambda$4959, 5007)/H$\beta$.  Many other approaches
or calibrations have been investigated \citep[see][and references
  within]{pilyugin16, peimbert17}.

\citet{shaver83} employed a novel approach by using RRLs to determine
the electron temperature instead of optical data and then applied
these values of $T_{\rm e}$ to the empirical formulae to calculate the
O/H abundance ratio.  The \citet{shaver83} metallicity-electron
temperature relationship therefore self-consistently uses the radio
data.  There have since been several studies, however, that produce
different results.  For example, \citet{pilyugin03} used the $T_{\rm
  e}$-method and found systematically lower O/H abundance ratios by
0.2--0.3 dex.  The different results are due to different atomic data
and different assumptions about the temperature structure.  For recent
studies with more sensitive observations and updated atomic data see
\citet{arellano-cordova20, arellano-cordova21}.

\begin{figure}
  \centering
  \includegraphics[angle=0,scale=0.75]{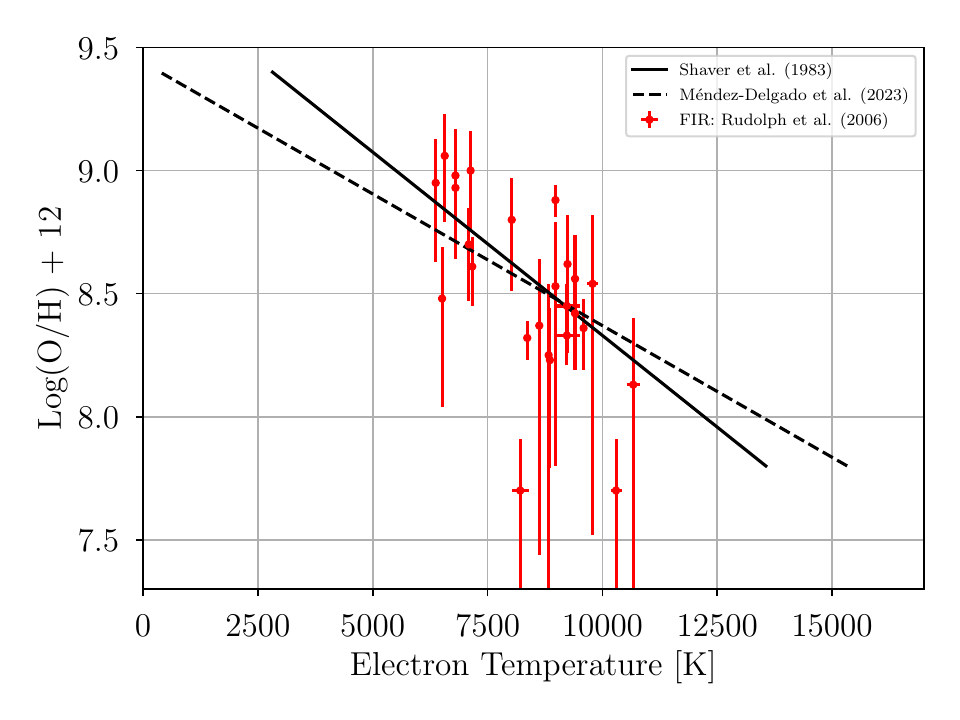}
  \caption{Metallicity-electron temperature \hii\ region relationships
    from \citet{shaver83} (solid line) and \citet{mendez-delgado23}
    (dashed line).  The red circles denote O/H abundance ratios from
    FIR CELs \citep{rudolph06} and electron temperatures from RRLs
    \citep{wenger19b}.  Here, we exclude the results from
    \citet{peeters02} which yield systematically lower O/H abundance
    ratios.}
  \label{fig:terrl-o2hfir}
\end{figure}

Temperature fluctuations within the nebula can produce different
evaluations of $T_{\rm e}$ depending on the method and therefore
different O/H abundance ratios \citep{peimbert67}.  This is thought to
be at least one explanation for the discrepancy between heavy element
abundances derived from CELs and optical recombination lines (ORLs)
that has existed for many years \citep{wyse42}.
\citet{mendez-delgado23} suggested that temperature fluctuations are
confined to the central regions of a nebula, which primarily affects
the highly ionized gas traced by [\oiii], and causes the abundance
discrepancy problem.  They derived a new metallicity-electron
temperature relationship, based on data from both Galactic and
extragalacitc \hii\ regions, appropriate when $T_{\rm e}$ is
determined using recombination lines and therefore appropriate for
RRLs:
\begin{equation}
{\rm Log(O/H)} + 12 = (9.44 \pm 0.08) - (1.07 \pm 0.09)T_{\rm e}/10^{4}.
\label{eq:mendez-delgado}
\end{equation}
The intercept of this linear relationship is similar to that of
\citet{shaver83} but the slope is about 1.4 times smaller (see
Figure~\ref{fig:terrl-o2hfir}).  \citet{shaver83} estimated that
temperature fluctuations in their \hii\ region sample were small,
$t^{2} \lsim 0.015$, based on values of $T_{\rm e}$ derived from RRLs
and CELs.  Here, $t^{2}$ is the root-mean-square (rms) deviation from
the average electron temperature.  In contrast,
\citet{mendez-delgado23} estimate $t^{2} > 0.025$ for most of the
\hii\ regions in their sample.  The quality of the optical spectra and
the accuracy of the atomic data are clearly very different between the
two studies.  Moreover, the sample by \citet{mendez-delgado23}
includes giant \hii\ regions with lower densities and metallicities.
Therefore, a comparison between the two metallicity-electron
temperature relationships may not be appropriate.

Another approach is to use CELs at far-infrared (FIR) wavelengths.
They have several advantages over optical CELs: (1) there is less
extinction from dust; and (2) they are not very sensitive to electron
temperature and thus temperature fluctuations.  But FIR CELs are
sensitive to electron density and are less bright then their optical
counterparts.  \citet{rudolph06} reanalyzed observations from the
literature \citep{simpson95, afflerbach97, rudolph97, peeters02} in a
self-consistent way to derive the O/H abundance ratio toward
\hii\ regions in the Galactic disk using the FIR lines of [\oiii]
(52\micron\ and 88\micron).  Their results are plotted in
Figure~\ref{fig:terrl-o2hfir} where the electron temperatures have
been determined from RRLs \citep{wenger19b}.\footnote{The electron
  temperature uncertainties from \citet{wenger19b} are only
  statistical, the systematic effects discussed here will increases
  the size of these error bars.}  The large uncertainties in the
FIR-determined O/H abundance ratios cannot distinguish between the two
metallicity-electron temperature relationships.

Because other factors besides metallicity will affect the electron
temperature, no single linear relationship will hold between O/H and
$T_{\rm e}$.  Moreover, most \hii\ regions in the Milky Way are not
accessible to optical studies due to dust extinction
\citep{anderson14}, and therefore the diagnostic tracers of various
nebular properties (e.g., $t^{2}$) are not available.  Here, we
therefore use the CLODUY simulations to assess the uncertainties in
the metallicity-electron temperature relationship and provide
correction factors to Log(O/H) + 12 when using radio data.
Specifically, the correction factor is the ratio of Log(O/H) + 12
input into the CLOUDY simulation to the Log(O/H) + 12 value derived
from the Shaver et al. metallicity-electron temperature relationship
using the LTE electron temperature calculated from the synthetic
H87$\alpha$ RRL in the CLOUDY simulation.  The fractional uncertainty
is just the correction factor minus 1.0.

\begin{figure}
  \centering
  \includegraphics[angle=0,scale=0.55]{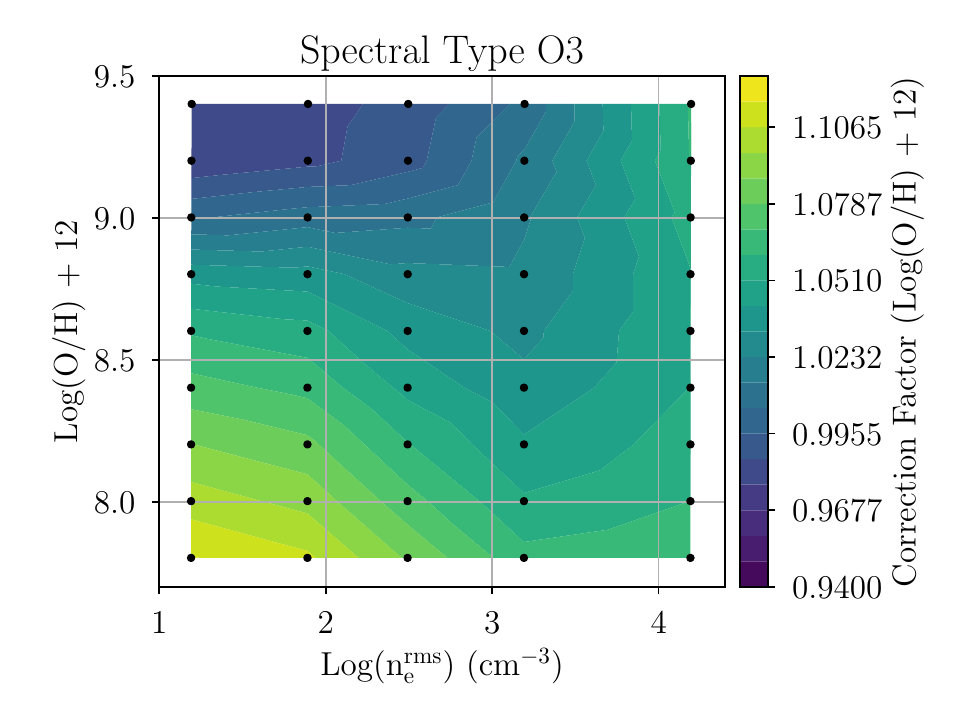}
  \includegraphics[angle=0,scale=0.55]{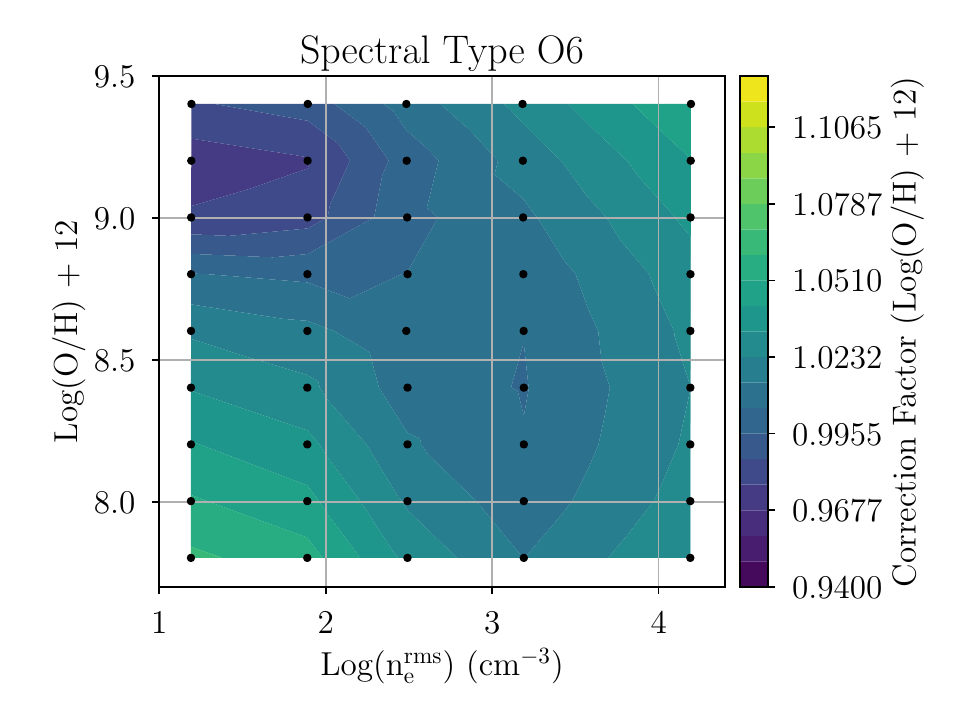}
  \includegraphics[angle=0,scale=0.55]{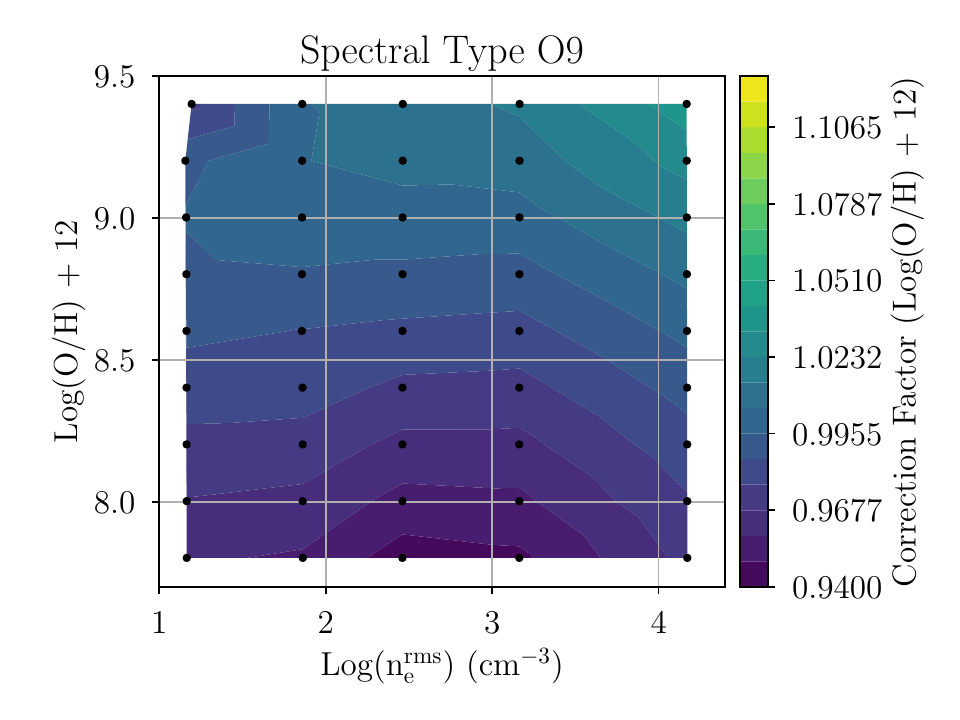}
  \caption{Correction factors of Log(O/H) + 12 for ionizing stars with
    spectral-types O3, O6, and O9.  The correction factor is the ratio
    of the O/H abundance input into the CLOUDY simulation to the O/H
    abundance derived from the \citet{shaver83} metallicity-electron
    temperature relationship using the LTE electron temperature
    calculated from the synthetic H87$\alpha$ RRL in the CLOUDY
    simulation.  The black points correspond to the location, defined
    by Log(O/H) + 12 and $n_{\rm e}^{\rm rms}$ from each simulation,
    from which the contours are generated.  The contour levels are
    between 0.94 and 1.125 with an increment of 0.00925.  For lower
    metallicities the correction factor depends on the spectral-type
    of the ionizing star, whereas for higher metallicities the
    correction factor is sensitive to the electron density.}
  \label{fig:cfo2h}
\end{figure}

The results are summarized as contour plots in Figure~\ref{fig:cfo2h}
for spectral types O3, O6, and O9.  The correction factors depend on
the O/H abundance ratios input into the CLOUDY simulation and the
electron density.  Here we derive a rms electron density in the same
way as would be done using radio observations: $n_{\rm e}^{\rm rms} =
\sqrt{EM/D}$.  For the CLOUDY simulations we calculate the emission
measure as $EM = \sum_{\rm i=1}^{\rm N} n_{\rm e,i}^{2}
\Delta\ell_{\rm i}$, where $n_{\rm e,i}$ is the electron density in
numerical zone i, $\Delta\ell_{\rm i}$ is the width of zone i, and N
is the number of zones.

Table~\ref{tab:cf} summarizes the correction factors for both the
electron temperature and O/H abundance ratio.  Listed are the spectral
type of the ionizing star, the rms electron density, the metallicity
given by Log(O/H) + 12, and the correction factors.  The correction
factors are a measure of the uncertainty when using radio tracers to
determine metallicity, but in principle they may also be used to
correct for systematic uncertainties if the spectral type, electron
density, and metallicity can be estimated \citep[e.g.,][]{schraml69}.
This is an iterative process because the correction factor is a
function of metallicity.  Overall, the correction factors are less
than 10\% from unity.  For lower metallicities the correction factor
depends on the spectral-type of the ionizing star, whereas for higher
metallicities the correction factor is sensitive to the electron
density.

\section{Summary}\label{sec:summary}

Heavy element abundances, or the metallicity, in \hii\ regions provide
important constraints to Galactic chemical evolution models.  Radio
recombination lines from \hii\ regions are one of the few tracers that
are not affected by dust and therefore probe the entire Galactic disk.
RRL emission from elements heavier than helium, however, is too weak
to be detected in \hii\ regions.  Since metals act as coolants they
primarily regulate the thermal motions of the ionized gas to produce a
linear relationship between metallicity and electron temperature.
Assuming LTE, the ratio of the RRL to the radio free-free continuum
provides a measure of the electron temperature independent from the
electron density, and therefore a way to indirectly determine the
metallicity \citep{wenger19b}.

Here, we use CLOUDY simulations to investigate the uncertainties in
this indirect method of determining the metallicity from radio data.
We run 135 CLOUDY simulations varying the spectral type, electron
density, and metallicity (defined by the O/H abundance ratio).  We
find that electron temperatures derived assuming LTE are about 20\%
higher, but overall LTE is a good assumption for cm-wavelength RRLs.
Overall, the CLOUDY simulation results are consistent with the
metallicity-electron temperature relationship determined empirically
by \citet{shaver83}.  But there exists significant dispersion since
ionizing stars with earlier spectral types or nebulae with higher
electron density yield higher electron temperatures.  When combined
the errors in the predicted metallicity, defined by Log(O/H) + 12, are
less than 10\%.  We derive correction factors to the \citet{shaver83}
metallicity-electron temperature relationship that depend on the
spectral type, electron density, and metallicity.

\begin{acknowledgments}
  
D.S.B thanks Dick Henry for suggesting this project.  We thank Gary
Ferland, Fran Guzm{\'a}n, and Marios Chatzikos for their help in
modeling RRLs in CLOUDY.  The National Radio Astronomy Observatory is
a facility of the National Science Foundation operated under
cooperative agreement by Associated Universities, Inc.  This research
has made use of NASA’s Astrophysics Data System Bibliographic
Services.  T.V.W. is supported by a National Science Foundation
Astronomy and Astrophysics Postdoctoral Fellowship under award
AST2202340.

\end{acknowledgments}

\vspace{5mm}

\software{CLOUDY \citep{chatzikos23}, Astropy (Astropy Collaboration et
  al. 2013), Matplolib \citep{hunter07}, NumPY \& SciPy
  \citep{vanderwalt11}.}

\startlongtable 
\begin{deluxetable}{ccccc}
\tabletypesize{\small}
\tablecaption{Correction Factor \label{tab:cf}}
\tablewidth{0pt}
\tablehead{
\colhead{Spectral} & \colhead{Log($n_{\rm e}^{\rm rms}$)} & 
\colhead{Log(O/H) + 12} & \multicolumn{2}{c}{\underline{~~~~~Correction Factor~~~~~}} \\ 
\colhead{Type} & \colhead{(cm$^{-3}$)} & 
\colhead{} & \colhead{$T_{\rm e}$} & \colhead{Log(O/H) + 12}  
}
\startdata 
O3 & 1.191 & 8.4 & 0.7564 & 1.0730 \\ 
O3 & 1.890 & 8.4 & 0.7741 & 1.0668 \\ 
O3 & 2.492 & 8.4 & 0.8338 & 1.0483 \\ 
O3 & 3.191 & 8.4 & 0.9021 & 1.0349 \\ 
O3 & 4.191 & 8.4 & 0.9056 & 1.0511 \\ 
O3 & 1.191 & 8.2 & 0.7769 & 1.0882 \\ 
O3 & 1.890 & 8.2 & 0.7941 & 1.0811 \\ 
O3 & 2.492 & 8.2 & 0.8423 & 1.0607 \\ 
O3 & 3.191 & 8.2 & 0.9058 & 1.0432 \\ 
O3 & 4.191 & 8.2 & 0.9178 & 1.0550 \\ 
O3 & 1.191 & 8.0 & 0.7980 & 1.1019 \\ 
O3 & 1.890 & 8.0 & 0.8143 & 1.0942 \\ 
O3 & 2.492 & 8.0 & 0.8576 & 1.0731 \\ 
O3 & 3.191 & 8.0 & 0.9146 & 1.0523 \\ 
O3 & 4.191 & 8.0 & 0.9243 & 1.0603 \\ 
O3 & 1.191 & 7.8 & 0.8211 & 1.1163 \\ 
O3 & 1.890 & 7.8 & 0.8361 & 1.1083 \\ 
O3 & 2.492 & 7.8 & 0.8689 & 1.0868 \\ 
O3 & 3.191 & 7.8 & 0.9182 & 1.0633 \\ 
O3 & 4.191 & 7.8 & 0.9301 & 1.0681 \\ 
O3 & 1.191 & 8.6 & 0.7272 & 1.0592 \\ 
O3 & 1.890 & 8.6 & 0.7535 & 1.0543 \\ 
O3 & 2.492 & 8.6 & 0.8161 & 1.0386 \\ 
O3 & 3.191 & 8.6 & 0.8937 & 1.0300 \\ 
O3 & 4.191 & 8.6 & 0.8986 & 1.0505 \\ 
O3 & 1.191 & 8.8 & 0.6927 & 1.0383 \\ 
O3 & 1.890 & 8.8 & 0.7262 & 1.0362 \\ 
O3 & 2.492 & 8.8 & 0.8012 & 1.0262 \\ 
O3 & 3.191 & 8.8 & 0.8861 & 1.0250 \\ 
O3 & 4.191 & 8.8 & 0.8868 & 1.0506 \\ 
O3 & 1.192 & 9.0 & 0.6405 & 1.0037 \\ 
O3 & 1.891 & 9.0 & 0.6906 & 1.0094 \\ 
O3 & 2.493 & 9.0 & 0.7867 & 1.0113 \\ 
O3 & 3.192 & 9.0 & 0.8769 & 1.0221 \\ 
O3 & 4.192 & 9.0 & 0.8728 & 1.0548 \\ 
O3 & 1.193 & 9.2 & 0.5700 & 0.9785 \\ 
O3 & 1.892 & 9.2 & 0.6454 & 0.9836 \\ 
O3 & 2.494 & 9.2 & 0.7694 & 0.9916 \\ 
O3 & 3.193 & 9.2 & 0.8687 & 1.0157 \\ 
O3 & 4.193 & 9.2 & 0.8551 & 1.0607 \\ 
O3 & 1.195 & 9.4 & 0.5199 & 0.9796 \\ 
O3 & 1.894 & 9.4 & 0.6074 & 0.9827 \\ 
O3 & 2.496 & 9.4 & 0.7430 & 0.9892 \\ 
O3 & 3.195 & 9.4 & 0.8441 & 1.0070 \\ 
O3 & 4.195 & 9.4 & 0.8309 & 1.0613 \\ 
O6 & 1.190 & 8.4 & 0.7520 & 1.0320 \\ 
O6 & 1.889 & 8.4 & 0.7760 & 1.0252 \\ 
O6 & 2.491 & 8.4 & 0.8419 & 1.0095 \\ 
O6 & 3.190 & 8.4 & 0.9118 & 1.0042 \\ 
O6 & 4.190 & 8.4 & 0.9023 & 1.0232 \\ 
O6 & 1.190 & 8.2 & 0.7791 & 1.0423 \\ 
O6 & 1.889 & 8.2 & 0.7944 & 1.0349 \\ 
O6 & 2.491 & 8.2 & 0.8536 & 1.0152 \\ 
O6 & 3.190 & 8.2 & 0.9147 & 1.0054 \\ 
O6 & 4.190 & 8.2 & 0.9112 & 1.0246 \\ 
O6 & 1.190 & 8.0 & 0.8008 & 1.0521 \\ 
O6 & 1.889 & 8.0 & 0.8216 & 1.0444 \\ 
O6 & 2.491 & 8.0 & 0.8686 & 1.0219 \\ 
O6 & 3.190 & 8.0 & 0.9245 & 1.0085 \\ 
O6 & 4.190 & 8.0 & 0.9169 & 1.0274 \\ 
O6 & 1.190 & 7.8 & 0.8246 & 1.0623 \\ 
O6 & 1.889 & 7.8 & 0.8365 & 1.0547 \\ 
O6 & 2.491 & 7.8 & 0.8754 & 1.0302 \\ 
O6 & 3.190 & 7.8 & 0.9256 & 1.0139 \\ 
O6 & 4.190 & 7.8 & 0.9285 & 1.0324 \\ 
O6 & 1.191 & 8.6 & 0.7228 & 1.0218 \\ 
O6 & 1.890 & 8.6 & 0.7533 & 1.0164 \\ 
O6 & 2.484 & 8.6 & 0.8922 & 1.0074 \\ 
O6 & 3.189 & 8.6 & 0.9130 & 1.0049 \\ 
O6 & 4.191 & 8.6 & 0.8921 & 1.0253 \\ 
O6 & 1.191 & 8.8 & 0.6844 & 1.0051 \\ 
O6 & 1.890 & 8.8 & 0.7234 & 1.0028 \\ 
O6 & 2.492 & 8.8 & 0.8246 & 1.0049 \\ 
O6 & 3.187 & 8.8 & 0.9178 & 1.0074 \\ 
O6 & 4.191 & 8.8 & 0.8777 & 1.0284 \\ 
O6 & 1.191 & 9.0 & 0.6291 & 0.9783 \\ 
O6 & 1.890 & 9.0 & 0.6843 & 0.9823 \\ 
O6 & 2.489 & 9.0 & 0.8460 & 1.0021 \\ 
O6 & 3.186 & 9.0 & 0.9209 & 1.0121 \\ 
O6 & 4.192 & 9.0 & 0.8607 & 1.0345 \\ 
O6 & 1.192 & 9.2 & 0.5905 & 0.9718 \\ 
O6 & 1.891 & 9.2 & 0.6746 & 0.9762 \\ 
O6 & 2.486 & 9.2 & 0.9289 & 0.9998 \\ 
O6 & 3.184 & 9.2 & 0.9309 & 1.0179 \\ 
O6 & 4.193 & 9.2 & 0.8403 & 1.0414 \\ 
O6 & 1.193 & 9.4 & 0.6196 & 0.9852 \\ 
O6 & 1.892 & 9.4 & 0.7724 & 0.9905 \\ 
O6 & 2.484 & 9.4 & 0.9406 & 1.0090 \\ 
O6 & 3.182 & 9.4 & 0.9493 & 1.0260 \\ 
O6 & 4.194 & 9.4 & 0.8089 & 1.0500 \\ 
O9 & 1.164 & 8.4 & 0.8744 & 0.9817 \\ 
O9 & 1.860 & 8.4 & 0.9352 & 0.9812 \\ 
O9 & 2.461 & 8.4 & 1.0045 & 0.9749 \\ 
O9 & 3.163 & 8.4 & 1.0401 & 0.9740 \\ 
O9 & 4.171 & 8.4 & 0.9719 & 0.9897 \\ 
O9 & 1.164 & 8.2 & 0.8933 & 0.9744 \\ 
O9 & 1.861 & 8.2 & 0.9393 & 0.9733 \\ 
O9 & 2.461 & 8.2 & 1.0218 & 0.9653 \\ 
O9 & 3.163 & 8.2 & 1.0482 & 0.9652 \\ 
O9 & 4.171 & 8.2 & 0.9804 & 0.9824 \\ 
O9 & 1.165 & 8.0 & 0.9111 & 0.9673 \\ 
O9 & 1.862 & 8.0 & 0.9481 & 0.9653 \\ 
O9 & 2.460 & 8.0 & 1.0313 & 0.9554 \\ 
O9 & 3.163 & 8.0 & 1.0554 & 0.9565 \\ 
O9 & 4.172 & 8.0 & 0.9833 & 0.9760 \\ 
O9 & 1.165 & 7.8 & 0.9337 & 0.9598 \\ 
O9 & 1.862 & 7.8 & 0.9549 & 0.9573 \\ 
O9 & 2.460 & 7.8 & 1.0489 & 0.9449 \\ 
O9 & 3.162 & 7.8 & 1.0625 & 0.9474 \\ 
O9 & 4.172 & 7.8 & 0.9962 & 0.9704 \\ 
O9 & 1.163 & 8.6 & 0.8564 & 0.9882 \\ 
O9 & 1.857 & 8.6 & 0.9491 & 0.9860 \\ 
O9 & 2.461 & 8.6 & 0.9888 & 0.9843 \\ 
O9 & 3.163 & 8.6 & 1.0329 & 0.9830 \\ 
O9 & 4.170 & 8.6 & 0.9611 & 0.9979 \\ 
O9 & 1.163 & 8.8 & 0.8376 & 0.9935 \\ 
O9 & 1.857 & 8.8 & 0.9241 & 0.9947 \\ 
O9 & 2.461 & 8.8 & 0.9718 & 0.9934 \\ 
O9 & 3.163 & 8.8 & 1.0251 & 0.9923 \\ 
O9 & 4.170 & 8.8 & 0.9502 & 1.0070 \\ 
O9 & 1.161 & 9.0 & 0.8298 & 0.9961 \\ 
O9 & 1.858 & 9.0 & 0.8973 & 1.0014 \\ 
O9 & 2.462 & 9.0 & 0.9553 & 1.0015 \\ 
O9 & 3.164 & 9.0 & 1.0177 & 1.0012 \\ 
O9 & 4.169 & 9.0 & 0.9362 & 1.0166 \\ 
O9 & 1.156 & 9.2 & 0.8259 & 0.9933 \\ 
O9 & 1.858 & 9.2 & 0.8704 & 1.0045 \\ 
O9 & 2.462 & 9.2 & 0.9402 & 1.0073 \\ 
O9 & 3.164 & 9.2 & 1.0120 & 1.0093 \\ 
O9 & 4.169 & 9.2 & 0.9196 & 1.0267 \\ 
O9 & 1.194 & 9.4 & 0.6392 & 0.9747 \\ 
O9 & 1.859 & 9.4 & 0.8566 & 1.0042 \\ 
O9 & 2.463 & 9.4 & 0.9404 & 1.0096 \\ 
O9 & 3.165 & 9.4 & 1.0103 & 1.0154 \\ 
O9 & 4.168 & 9.4 & 0.9023 & 1.0375 \\ 
\enddata 
\end{deluxetable}

\clearpage

\appendix

\section{CLOUDY Simulation Parameters}\label{sec:sim}

Below are the inputs for one CLOUDY simulation that consisted of a
O6-type ionizing star and a model nebula with $n_{\rm H} = 200$\percc,
$D = 2.5$\pc, and ${\rm Log(O/H)} + 12 = 8.8$.

\begin{verbatim}

title hii_O6_em5_m0.2_starz 
# commands controlling continuum ========= 
table star atlas Z-1.5 38151.0 
q(h) 48.96 
# commands controlling geometry  ========= 
sphere 
radius 0.001 to 1.25 linear parsecs 
# commands for density & abundances ========= 
hden 200.0 linear 
abundances "HII.abn" no grains 
metals 0.2 log 
filling 1. 
# number of levels to use 
database h-like resolved levels 25 
database h-like collapsed levels 375 
database he-like resolved levels 20 
database he-like collapsed levels 380 
# collisional excitation data (default is lebedev) 
#database h-like collisions lebedev 
# other commands for details     ========= 
iterate 
# Allow for lower temperatures (default is Te=4000K) 
stop temperature 1000.0 linear 
# commands controlling output    ========= 
# set continuum frequencies 
set nFnu add 3.00299 cm 
set nFnu add 3.05299 cm 
set nFnu add 3.10299 cm 
set nFnu add 3.15887 cm 
set nFnu add 3.20887 cm 
set nFnu add 3.26716 cm 
set nFnu add 3.31716 cm 
set nFnu add 3.37791 cm 
set nFnu add 3.42791 cm 
set nFnu add 3.49113 cm 
set nFnu add 3.54113 cm 
set nFnu add 3.60686 cm 
set nFnu add 3.65686 cm 
set nFnu add 3.72511 cm 
set nFnu add 3.77511 cm 
# save details about calculation and model 
save performance "hii_O6_em5_m0.2_starz.per" 
save overview last "hii_O6_em5_m0.2_starz.ovr" 
save dr last "hii_O6_em5_m0.2_starz.dr" 
save incident continuum last "hii_O6_em5_m0.2_starz.inc" 
save continuum last "hii_O6_em5_m0.2_starz.con" units microns 
save transmitted continuum last "hii_O6_em5_m0.2_starz.trn"  
save line list "hii_O6_em5_m0.2_starz.linaC" "linelistOnlyCont.dat" last absolute 
save line list "hii_O6_em5_m0.2_starz.lineaL" "linelistNoCont.dat" last emergent absolute 
save hydrogen lines alpha last "hii_O6_em5_m0.2_starz.hlin" 
save species departure coefficients last "hii_O6_em5_m0.2_starz.dep" "H[:]" 
# hii_O6_em5_m0.2_starz.in 


\end{verbatim}

\clearpage

\bibliography{ms}

\end{document}